\documentclass[aps, prb,twocolumn,superscriptaddress, nofootinbib, showpacs]{revtex4-1}

\usepackage{graphicx}
\usepackage{dcolumn}
\usepackage{dcolumn}
\usepackage{textgreek}
\usepackage{textcomp}
\usepackage{adjustbox}
\usepackage{multirow}

\def\genPyrochlore{\textit{RE}$_2$\textit{B}$_2$O$_7$}
\def\EGO{Er$_2$Ge$_2$O$_7$}
\def\HGO{Ho$_2$Ge$_2$O$_7$}

\def\DGO{Dy$_2$Ge$_2$O$_7$}
\def\DTO{Dy$_2$Ti$_2$O$_7$}
\def\HTO{Ho$_2$Ti$_2$O$_7$}
\def\ETO{Er$_2$Ti$_2$O$_7$}

\def\a{\textit{a}}

\def\b{\textit{b}}
\def\c{\textit{c}}

\def\coa{$c/a$}

\def\Tn{$T_N$}

\def\Na+{Na\textsuperscript{+}}
\def\Ba+{Ba\textsuperscript{2+}}

\def\degrees{$^{\circ}$}
\def\iA{\text{\AA}\textsuperscript{-1}}
\def\Asq{\text{\AA}\textsuperscript{2}}

\def\Pftt{$P4_12_12$}
\def\Gtw{$\Gamma_3$}

\begin{document}

\preprint{APS/123-QED}

\title{Local-Ising type magnetic order and metamagnetism in the rare-earth pyrogermanate Er$_2$Ge$_2$O$_7$}

\author{K.M. Taddei}
\email[corresponding author ]{taddeikm@ornl.gov}
\affiliation{Neutron Scattering Division, Oak Ridge National Laboratory, Oak Ridge, TN 37831}
\author{L. Sanjeewa}
\affiliation{Materials Science and Technology Division, Oak Ridge National Laboratory, Oak Ridge, TN 37831}
\affiliation{Department of Chemistry, Clemson University, Clemson, SC 29634 }
\author{J.W. Kolis}
\affiliation{Department of Chemistry, Clemson University, Clemson, SC 29634 }
\author{A.S. Sefat}
\affiliation{Materials Science and Technology Division, Oak Ridge National Laboratory, Oak Ridge, TN 37831}
\author{C. de la Cruz}
\affiliation{Neutron Scattering Division, Oak Ridge National Laboratory, Oak Ridge, TN 37831}
\author{D.M. Pajerowski}
\affiliation{Neutron Scattering Division, Oak Ridge National Laboratory, Oak Ridge, TN 37831}

\date{\today}

\begin{abstract}
 
The recent discoveries of proximate quantum spin-liquid compounds and their potential application in quantum computing informs the search for new candidate materials for quantum spin-ice and spin-liquid physics. While the majority of such work has centered on members of the pyrochlore family due to their inherently frustrated linked tetrahedral structure, the rare-earth pyrogermanates also show promise for possible frustrated magnetic behavior. With the familiar stoichiometry \textit{RE}$_2$Ge$_2$O$_7$, these compounds generally have tetragonal symmetry with a rare-earth  sublattice built of a spiral of alternating edge and corner sharing rare-earth site triangles. Studies on \DGO\ and \HGO\ have shown tunable low temperature antiferromagnetic order, a high frustration index and spin-ice like dynamics. Here we use neutron diffraction to study magnetic order in \EGO\ (space group \Pftt ) and find the lowest yet Ne\'el temperature in the pyrogermanates of 1.15 K. Using neutron powder diffraction we find the magnetic structure to order with $k = (0,0,0)$ ordering vector, magnetic space group symmetry $P4_{1}^{'}2_{1}2^{'}$ and a refined Er moment of $m = 8.1 \mu_B$ - near the expected value for the Er$^{3+}$ free ion. Provocatively, the magnetic structure exhibits similar \rq local Ising\lq\ behavior to that seen in the pyrocholres where the Er moment points up or down along the short Er-Er bond. Upon applying a magnetic field we find a first order metamagnetic transition at $\sim$ 0.35 T to a lower symmetry $P2_{1}^{'}2_{1}^{'}2$ structure. This magnetic transition involves an inversion of Er moments aligned antiparallel to the applied field describing a class I spin-flip type transition, indicating a strong local anisotropy at the Er site - reminiscent of that seen in the spin-ice pyrochlores.

\end{abstract}

\pacs{74.25.Dw, 74.62.Dh, 74.70.Xa, 61.05.fm}

\maketitle


\section{\label{sec:intro}Introduction}

Magnetically frustrated materials have long drawn enthusiam from the condensed matter community for their ability to strain familiar approximations, reveal new physics and host exotic ground states\cite{Wannier1950, Derrida1978,Ramirez1994}. This particular combination of attributes leads frustrated magnetic materials, at times, to present an accessible interface between condensed matter theory and experiment, where relatively simple, or at least enumerable, Hamiltonians are able to describe the rich physics observed \cite{Gardner2010, Derrida1978, Wannier1950}. Indeed, an early test system for commercially produced quantum computers has been frustrated spin glasses for this reason \cite{Harris2018}. This interface has been reinvigorated recently with the discovery of proximate quantum spin-liquid materials and the ostensible vitality of frustrated quantum magnetic states for quantum computing, bringing anew interest in the discovery of new frustrated materials \cite{Kitaev2003,Banerjee2016,Banerjee2017,Ma2017,Zhang2018,Zheng2017,Do2017,Balz2016}. 

Historically one of the most fruitfully studied families of frustrated magnetic materials has been the rare-earth pyrochlores (\genPyrochlore\ with \textit{RE} = La - Yb, \textit{B} = Ti, Ge, Sn) \cite{Harris1997, Fukazawa2002, Castelnovo2008, Gardner2010, Siddharthan1999}. With a structural motif of corner-sharing \textit{RE}-Ge tetrahedra which naturally gives rise to competing exchange interactions, these systems tend to complex frustrated magnetic ground states such as spin ices, glasses and liquids and consequently reveal emergent novel physics \cite{Bramwell1998,Bramwell2001,Greedan2006}. In members such as \DTO\ and \HTO , the local ion anisotropy forces the \textit{RE} site magnetic moment to point along the \textit{RE}-Ge tetrahedron's local $<111>$ direction \cite{Fukazawa2002, Bramwell2001}. This defines a local form of the Ising model where each \textit{RE} site can point parallel or antiparallel to the local $<111>$ axis mapping to the spin-up/spin-down Ising description \cite{Fukazawa2002}.  When the ferromagnetic (FM) exchange interaction between neighboring \textit{RE} sites is considered the famous \lq spin-ice rule\rq\ state is achieved which describes a strong frustration that confounds long range magnetic order down to the lowest measured temperatures of several mK \cite{Fukazawa2002, Krey2012}.

In the \textit{RE} pyrochlores, the large magnetic moment of the \textit{RE} site ensures that the exchange and single-ion anisotropy are not the only meaningful terms in the magnetic Hamiltonian - dipole-dipole interactions are, for instance, of relevant energy scales. Therefore, tuning the relative strength of the different magnetic interactions - and consequently the level of frustration - is possible and easily achived through changing the \textit{RE} ion (and consequently the magnetic moment size and crystal field levels) or the application of an external magnetic field \cite{Hiroi2003,Cao2009}. Such effects have been systematically studied and result in discrete changes in appropriate magnetic interaction models between neighboring \textit{RE} ions (such as Ising for \DTO\ and \textit{XY} for \ETO ) and even low field magnetic transitions under applied field (metamagnetic transitions) \cite{Cao2008a,Cao2008,Cao2009, Hiroi2003,Isakov2004}. In this way a phase diagram can be created where tuning the relative strengths of the exchange and dipole-dipole interactions leads to a bevy of possible magnetic ground states with exotic states accessible from neighboring relatively pedestrian ones via parameters easy to control with experimental and synthesis conditions \cite{Hertog2000}.

The interest and success of the pyrochlores in revealing new physics informs a search for similar materials whose properties might beget similar competing magnetic interactions. The \textit{RE} pyrogermanates (REPG) share the pyrochlore’s stoichiometry but with a lower symmetry nuclear structure (as low as space group $P\overline{1}$). Nonetheless these systems also exhibit an inherently geometrically frustrated \textit{RE} structural motif built of a spiral structure with alternating corner sharing and edge sharing \textit{RE} triangles \cite{Becker1987}.

Despite this potential, little work has been performed on the REPG family. Beyond the intial synthesis report, only the \HGO , \DGO\ and \EGO\ members of the REPG family have recieved further study \cite{Morosan2008,Ke2008, Ghosh1998}. These members crystallize with the tetragonal space group \Pftt\ and exhibit highly anisotropic magnetic susceptibilities.  In \HGO\ (for which neutron diffraction data has been reported), below 1.6 K magnetic ordering is seen with a large magnetic moment of 8$\mu_B$/Ho and complex magnetic structure with the Ho moments locked in the crystallographic \textit{ab} plane and rotating along the \textit{c}-axis \cite{Morosan2008}. Interestingly, in both the \DGO\ and \HGO\ materials, field dependent AC susceptibility measurements suggest the rare-earth ions behave like Ising-spins with spin-relaxation phenomena which indicate similar magnetic behavior to the spin-ice pyrochlores \cite{Cao2009, Bramwell1998,Snyder2001,Snyder2004,Ehlers2004}.

In this paper, we report neutron scattering studies on the \EGO\ REPG down to mK temperatures. Our work finds magnetic order below 1.15 K with a three-dimensional spiral structure unlike the co-planer structure reported in \HGO . This ordering temperature is well below that predicted from Curie-Weiss fitting of the high temperature magnetic susceptibility suggesting significant frustration. Upon application of a small ($ < 1$ T) external magnetic field, we find \EGO\ undergoes a metamagnetic transition. Single crystal neutron diffraction reveals that with the field applied in the easy plane of the \HGO\ material, the metamagnetic transition is of spin-flip type with moments anti-parallel to the applied field inverting through their crystallographic site. These results suggest a similarity between the \textit{RE}PG and the \textit{RE} pyrochlores with a similar geometric frustration and strong competition between different magnetic interaction mechanisms.   

\section{\label{sec:methods} Experimental Methods}

\subsection{\label{subsec:synthesis} Synthesis}

Single crystals of \EGO\ were synthesized using a direct combination of Er$_2$O$_3$ and GeO$_2$ via high temperature and high pressure hydrothermal synthesis. In a typical reaction 0.4 g of total reactants (0.2981 g of HEFA rare Earth 99.99\% Er$_3$O$_2$ and 0.1019 g of Alfa Aesar 99.9\% GeO$_2$) were used in 4:5 stoichiometric ratio. The crystalline products were grown at 650 \degrees C for 14 days in fine silver (99.9\% ) $3/8" \times\ 2.5"$ ampules loaded into a Tuttle cold seal autoclave constructed from Inconel 718 material. The ampules were loaded with the appropriate component oxides and weld sealed from both ends after addition of 0.8 mL of 20 M CsF as a mineralizer. Upon reaction completion, the silver ampules were opened and washed with deionized water. The yield of \EGO\ single crystals was $\sim 90$\% with the remainder being unreacted powder. Single crystal of \EGO\ were produced as pink plate shaped crystals approximately $1 \times\ 1 \times\ 0.5$ mm in size.   

Powder samples of \EGO\ were synthesized using a conventional solid state method. A mixture of total mass 5 g ( 3.2323 g Er$_2$O$_3$ and 1.7677 g GeO$_2$)was used with a stoichiometric ratio of 1:2. The reactants were mixed, ground and heat to 1000 \degrees C for 12 hrs. The resultant powders were pressed into pellets and calcined at 1250 \degrees C for 1 day. To ensure homogeneity, several reheating and regrinding steps were performed until no further impurities were present in the sample.

Initial characterization of the single crystal and powder samples were carried out via room temperature single crystal and powder x-ray diffraction using a Bruker D8 Venture with Incoatec Mo K$_\alpha$ microfocus source Photon 100 CMOS detector and a Rigaku Ultima IV diffractometer with CuK$_\alpha$ radiation respectively. The collected single crystal data was processed and scaled using the Apex3 (SAINT and SADABS) software suites and Rietveld refinements were performed using the SHELXTL software suite \cite{Sheldrick2015}.

Magnetization measurements were carried out using a magnetic property measurement
system (Quantum Design) using finely ground crystals of \EGO . Temperature dependent data were collected upon warming in a magnetic field of 10 kOe. Isothermal magnetization versus applied field curves were collected at 2 K.

\subsection{\label{subsec:neutron} Neutron Scattering Experiments}

Neutron powder diffraction measurements were performed on the HB-2A (POWDER) beamline of Oak Ridge National Laboratory's High Flux Isotope Reactor. Approximately 4.5 g of powder \EGO\ were placed into an aluminum powder can sealed under Helium atmosphere. Field and temperature dependent measurements were performed using a cryomagnet loaded with a 3He cryostick, allowing for field and temperature ranges of 0 - 5 T and 0.5 - 300 K respectively. 

Diffraction patterns were collected on HB-2A using the open-21'-12' collimator settings (for pre-monochromator, pre-sample and pre-detector collimation respectively) with both the short wavelength, high-intensity 1.54\AA\ and longer wavelength \lq magnetism-optimized\rq\ 2.41\AA\ monochromator reflections. Powder patterns were collected over a scattering vector ($Q$) range of  0.09 \iA\ $< Q < $ 4.63 \iA\ with count times between 1 and 8 hours per scan. For temperature dependent scans of peak intensities, a detector was positioned to be centered in $2\theta$ on the peak position and then the temperature was increased as scattering rates counts were taken using counting times of 200 sec/temperature.  

Analysis of the neutron powder diffraction data was performed using the Rietveld method as implemented in the FullProf software suite.\cite{Rodriguez-Carvajal1993} The Thompson-Cox-Hasting formulation for a psuedo-Voight peak shape with axial divergence asymmetry was used to fit the instrumental profile of HB-2A \cite{Finger1994}. In addition to profile fitting, the atomic positions, atomic displacement parameters of all sites as well as the fractional occupancies were refined. For magnetic structure determination the Simulated Annealing and Representational Analysis (SARAh) and ISODISTORT software programs were used \cite{Wills2000, Campbell2006}. Visualization of the crystal structure was performed using VESTA \cite{Momma2011}.

Single crystal neutron diffraction measurements were performed on the HB-1 and HB-3 triple-axis beamlines of HFIR with use of cryomagnet loaded with a 3He insert. Due to the small size of the as grown crystals, three crystals were coaligned for measurement totaling in 3.8 mg. The crystals were mounted on an aluminum pin with a small amount of varnish and the pin was masked with Gd$_2$O$_3$ to minimize contamination from scattering off the aluminum mount. The alignment of the crystals and the geometry of the triple axis spectrometers allowed access to the $H0L$ scattering plane. To maximize flux, loose collimation was used with the $\lambda = 2.36$\AA\ 14.7 meV incident beam.  

\section{\label{sec:results} Results and Discussion}

\subsection{\label{subsec:rt} Structure}

\begin{figure}
	\includegraphics[width=\columnwidth]{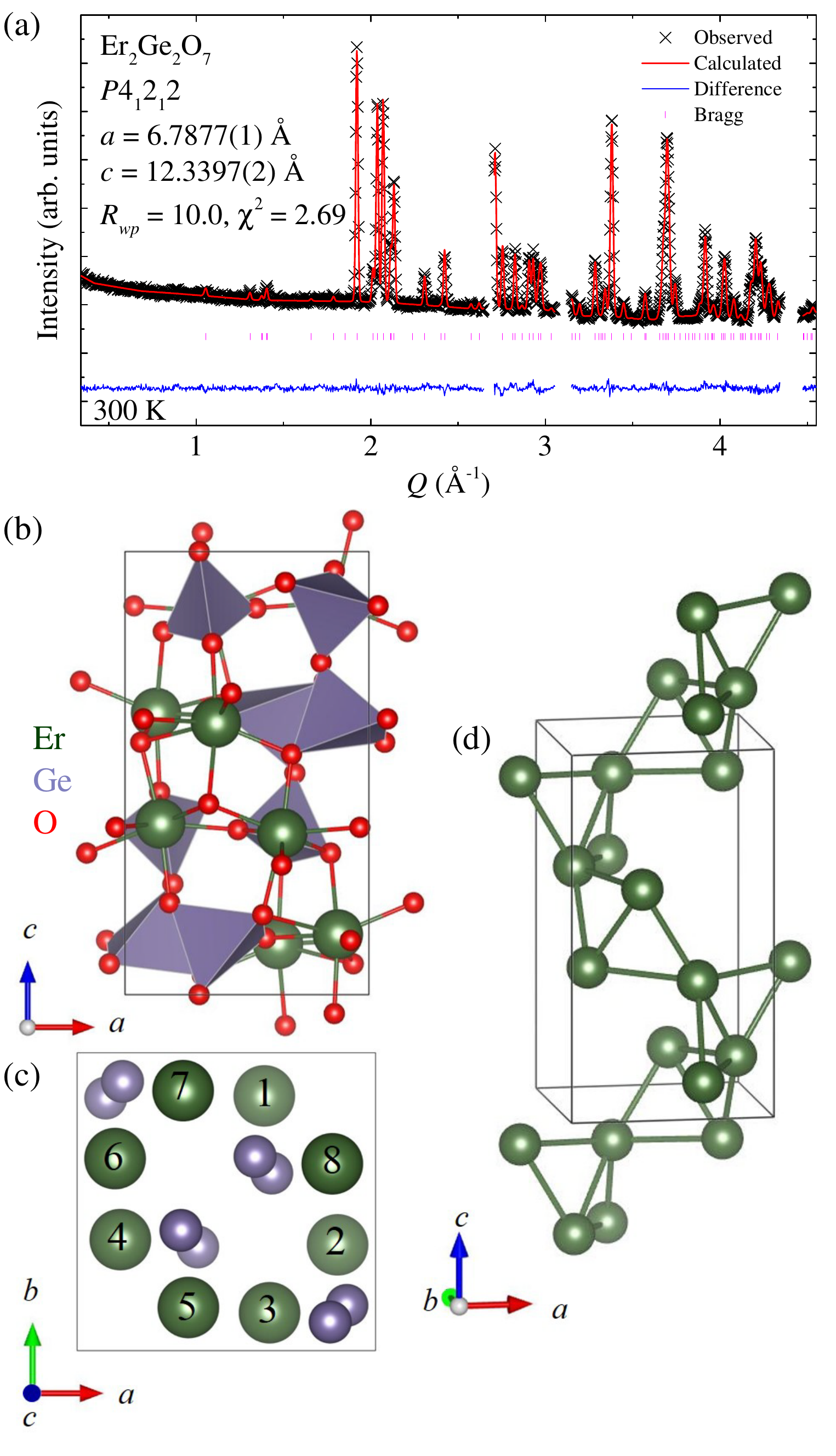}
	\caption{\label{fig:one} (a) Neutron diffraction pattern and Rietveld refinement of \EGO\ with space group symmetry \Pftt , data collected at 300 K with an incident beam of 2.41 \AA . (b) \Pftt\ structure with Er, Ge and O atoms in green, purple and red respectively . (c) View of unit cell along the \c -axis with O atoms removed for clarity. Er sites are numbered in sets per layer so that atoms 1 and 2 describe the bottom layer and so on - these labels are not consistent with the atom labels from symmetry operations. (d) Er sublattice highlighting the alternating edge and corner sharing sharing Er$_3$ triangle unit (e). Gaps in the diffraction pattern shown in panel (a) are regions containing Al peaks from the sample can which have been excluded from the refinement.}	
\end{figure}

\EGO\ is reported to crystallize with the tetragonal non-centrosymmetric \Pftt\ space group symmetry \cite{Becker1987}. Our studies corroborate these results as seen in Figure~\ref{fig:one}(a) which shows a neutron powder diffraction pattern of \EGO\ collected under ambient conditions modeled with the reported \Pftt\ structure. The model produces a satisfactory agreement with the data with $R_{wp}$ and $\chi^{2}$ parameters of 10.0 and 2.69 respectively. We note that while visually the fit looks quite good, we obtain a larger than expected $R_{wp}$. We attribute this to the significant region of the fit ($Q < 2$ \iA) which consists mainly of a large sloping background. This feature of the data will be discussed more later. The obtained refinement parameters for the 300, 2 and 0.5 K data are reported in Table~\ref{tab:one}.

The refined structure of \EGO\ is shown in Figure~\ref{fig:one}(b-d). As enumerated in Table~\ref{tab:one}, the \Pftt\ structure has one independent crystallographic site each for Er and Ge but four independent sites for the O atoms three of whose site symmetry places no restrictions on the \textit{x,y,z} positions. This leads to significantly different \textit{A} and \textit{B} site O coordination than in the pyrochlores. In the \textit{RE}PG, the \textit{A}-site is coordinated by seven surrounding oxygen rather than eight creating a highly distorted pentagonal bi-pyramid rather than the distorted cube (Fig.~\ref{fig:one}(b)). Due to the low symmetry of the O sites, the bonding within the ErO$_7$ bi-pyraminds is highly variable ranging from $\sim 2.21$ to $\sim 2.37$ \AA .  The pentagonal bi-pyramids are edge sharing creating a continuous Er helix along the \c -direction. The \textit{B}-site Ge atoms are tetrahedrally coordinated by O and form corner-sharing tetrahedron (Fig.~\ref{fig:one}(b)) with bond lengths between $\sim 1.78$ and $\sim 1.74$ \AA .

\begin{table}
	\caption{\label{tab:one}Crystallographic parameters of \EGO\ at 300, 2, and 0.5 K. Parameters determined from Rietveld refinements performed using the 300, 2 and 0.5 K data collected with the 1.54 \iA\ wavelength. The atomic displacement parameters and magnetic moment are reported in units of \Asq\ and $\mu _B/Er$ respectively. }
	\begin{ruledtabular}
		\begin{tabular}{llll}
     		 \multicolumn{1}{c}{ } & \multicolumn{1}{c}{300 K} & \multicolumn{1}{c}{2.0 K} & \multicolumn{1}{c}{0.5 K} \\ 
	\hline
	\multicolumn{1}{l}{Space Group} & \multicolumn{1}{c}{\Pftt} & \multicolumn{1}{c}{\Pftt} & \multicolumn{1}{c}{\Pftt} \\
	\multicolumn{1}{l}{$R_{wp}$} & \multicolumn{1}{c}{7.17} & \multicolumn{1}{c}{9.40} & \multicolumn{1}{c}{8.29} \\
	\multicolumn{1}{l}{$\chi^2$} & \multicolumn{1}{c}{1.96} & \multicolumn{1}{c}{3.79} & \multicolumn{1}{c}{3.10} \\
	\multicolumn{1}{l}{$a$ (\AA)} & \multicolumn{1}{c}{6.7877(1)} & \multicolumn{1}{c}{6.7829(1)} & \multicolumn{1}{c}{6.7826(1)} \\
	\multicolumn{1}{l}{$c$ (\AA)} & \multicolumn{1}{c}{12.3397(2)} & \multicolumn{1}{c}{12.3319(2)} & \multicolumn{1}{c}{12.3317(3)} \\
	\multicolumn{1}{l}{$c/a$} & \multicolumn{1}{c}{1.8180(2)} & \multicolumn{1}{c}{1.8181(2)} & \multicolumn{1}{c}{1.8181(2)} \\
		\multicolumn{1}{l}{$V$(\AA$^3$)} & \multicolumn{1}{c}{568.52(1)} &
	\multicolumn{1}{c}{567.36(3)} & \multicolumn{1}{c}{567.30(3)} \\
	\hline
\multicolumn{1}{l}{Er ($8a$)}	&		&		&		\\
\multicolumn{1}{r}{$x$}	&	0.8770(3)	&	0.8741(4)	&	0.8743(4)	\\
\multicolumn{1}{r}{$y$}	&	0.3553(3)	&	0.3547(5)	&	0.3545(4)	\\
\multicolumn{1}{r}{$z$}	&	0.1354(2)	&	0.1358(2)	&	0.1360(2)	\\
\multicolumn{1}{r}{$U$}	&	0.0054(6)	&	0.0027(9)	&	0.00060(8)	\\
\multicolumn{1}{r}{$M$}	&		&		&	8.1(3)	\\
\multicolumn{1}{l}{Ge ($8a$)} 	&		&		&		\\
\multicolumn{1}{r}{$x$} 	&	0.9014(3)	&	0.9001(5)	&	0.9008(6)	\\
\multicolumn{1}{r}{$y$} 	&	0.1534(3)	&	0.1514(5)	&	0.1508(6)	\\
\multicolumn{1}{r}{$z$} 	&	0.6197(2)	&	0.6188(3)	&	0.6181(3)	\\
\multicolumn{1}{r}{$U$} 	&	0.0068(5)	&	0.0060(9)	&	0.0091(1)	\\
\multicolumn{1}{l}{O1 ($4a$)} 	&		&		&		\\
\multicolumn{1}{r}{$x$}	&	0.8045(4)	&	0.8038(6)	&	0.8028(8)	\\
\multicolumn{1}{r}{$y$}	&	0.1956(4)	&	0.1962(6)	&	0.1972(8)	\\
\multicolumn{1}{r}{$z$}	&	0.75	&	0.75	&	0.75	\\
\multicolumn{1}{r}{$U$} 	&	0.013(1)	&	0.0065(2)	&	0.015(3)	\\
\multicolumn{1}{l}{O2 ($8a$)} 	&		&		&		\\
\multicolumn{1}{r}{$x$}	&	0.0786(5)	&	0.0769(7)	&	0.0744(8)	\\
\multicolumn{1}{r}{$y$}	&	-0.0327(4)	&	-0.0321(6)	&	-0.0318(7)	\\
\multicolumn{1}{r}{$z$}	&	0.6233(3)	&	0.6242(4)	&	0.6247(5)	\\
\multicolumn{1}{r}{$U$}	&	0.0072(8)	&	0.0036(1)	&	0.0039(1)	\\
\multicolumn{1}{l}{O3 ($8a$)} 	&		&      	&		\\
\multicolumn{1}{r}{$x$}	&	0.0639(4)	&	0.0648(6)	&	0.0663(8)	\\
\multicolumn{1}{r}{$y$}	&	0.3399(6)	&	0.3379(9)	&	0.337(1)	\\
\multicolumn{1}{r}{$z$}	&	0.5710(3)	&	0.5709(4)	&	0.5719(5)	\\
\multicolumn{1}{r}{$U$}	&	0.010(9)	&	0.009(1)	&	0.01(2)	\\
\multicolumn{1}{l}{O4 ($8a$)} 	&		&		&		\\
\multicolumn{1}{r}{$x$}	&	0.6828(4)	&	0.6844(6)	&	0.6839(8)	\\
\multicolumn{1}{r}{$y$}	&	0.1409(5)	&	0.1439(7)	&	0.1458(9)	\\
\multicolumn{1}{r}{$z$}	&	0.5436(2)	&	0.5449(4)	&	0.5456(4)	\\
\multicolumn{1}{r}{$U$}	&	0.010(8)	&	0.0054(1)	&	0.0049(1)	\\

		\end{tabular}
	\end{ruledtabular}
\end{table}

Along the tetragonal axis, the Er sublattice can be divided into four layers each with two Er which stack to create a spiral, wraping around the \c -axis (Fig.~\ref{fig:one}(b) and (c)). In analogy to the pyrochlores, we can also consider how the Er are self coordinated. As shown in Figure~\ref{fig:one}(d), the Er form a unit of edge sharing triangles which are bent along the short axis. These units are then linked through corner sharing alternately along their short and long axes creating the Er sublattice helix. This structure was also well-described in Ref.~\onlinecite{Morosan2008} for the \HGO\ \textit{RE}PG member.

Compared to \HGO\ structurally, we find \EGO\ to have slightly reduced \a\ and \c\ lattice parameters with a reduction of $\sim$ 0.3 \%\ in each direction. This contraction of the unit cell is consitent with the slightly smaller ionic radius of Er$^{3+}$ and results in a slight contraction of the in-plane and out-of-plane \textit{RE}-\textit{RE} distances as well as the \textit{RE}-O bond lengths \cite{Shannon1976}. However, no significant broader changes are seen in either the Er sublattice or in the ErO$_7$ polyhedron. As will be discussed in Section~\ref{subsec:mag}, the magnetic interactions in the \textit{RE}PG are expected to be complex with competing interactions. This 0.3\%\ contraction then gives a possible tuning parameter - or additional effect to consider - in the determination of magnetic order. Here, the \textit{RE} sites are brought closer together which naively should increase dipole-type interactions between the anticipated large magnetic moments on these sties. 

Diffraction patterns were collected at 300, 2, 1.2, 0.8 and 0.5 K allowing minimal tracking of the lattice parameters' temperature dependence. As shown in Table~\ref{tab:one} between 2 and 300 K both the \a\ and \c\ lattice parameters dilate, as expected for thermal expansion. The measure of the expansion's anisotropy is obtained by taking the ratio $c/a$. From 300 to 2 K the \coa\ ratio shows no change within the certainty of our measurements indicating an isotropic contraction.

\subsection{\label{subsec:mag} Zero-Field Magnetic Structure}

\begin{figure}
	\includegraphics[width=\columnwidth]{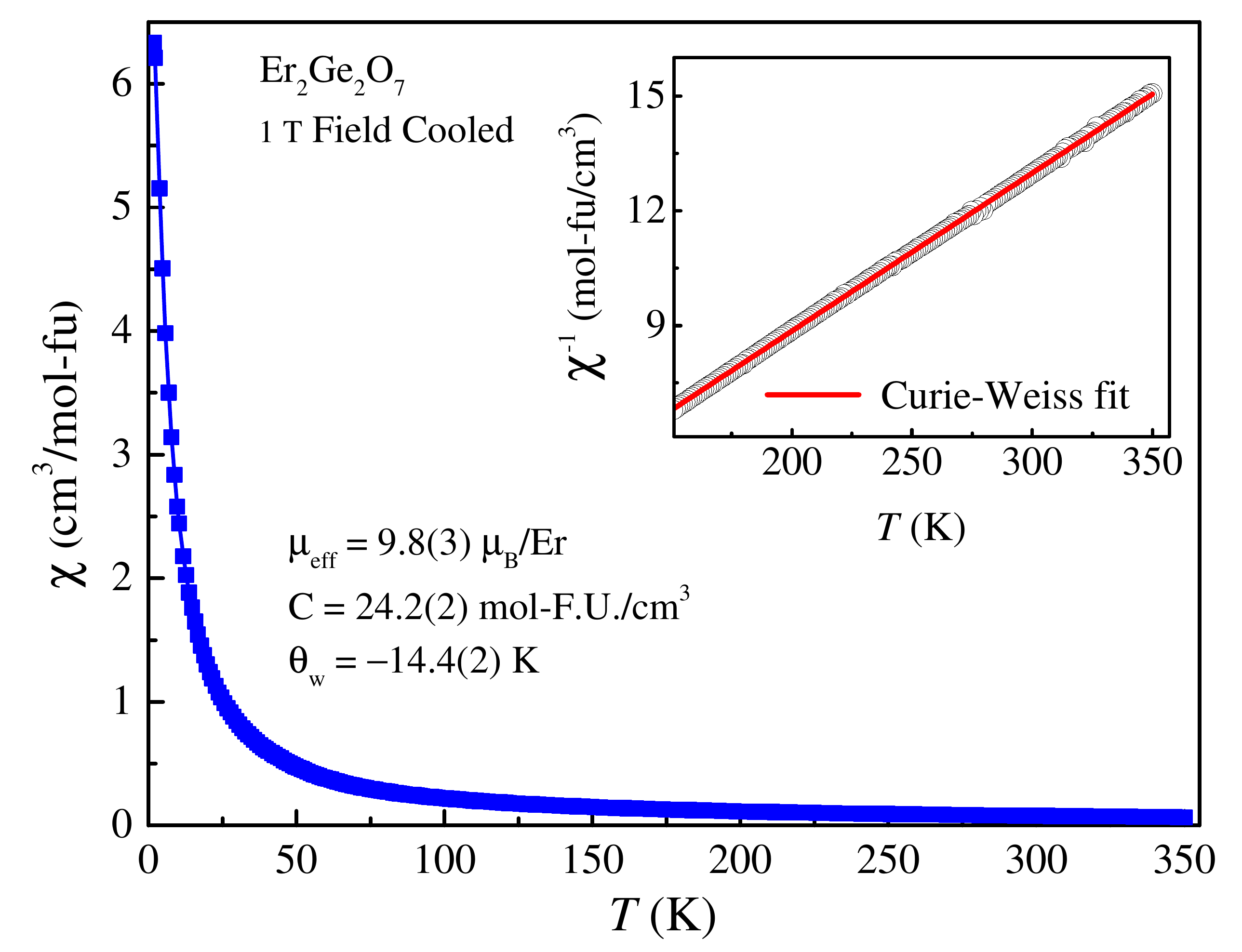}
	\caption{\label{fig:two} Susceptibility measurements performed on powder sample of \EGO\ under a 1 T field following a field-cooling procedure. Fitting of the inverse susceptibility shown in the inset was performed using the standard Curie-Weiss law $\chi ^{-1} = C(T-\theta_{w})$. }	
\end{figure}

While previous reports on the \DGO\ and \HGO\ \textit{RE}PG found evidence of magnetic order in transport measurements and solved the magnetic structure using neutron diffraction respectively, no magnetic order has yet been reported in \EGO\ \cite{Morosan2008, Ke2008}. Early work measuring the anisotropic susceptibility suggested possible order with a Curie-Weiss temperature ($\theta_{w}$) of $\sim$ 6 K \cite{Ghosh1998}. However, no report, the authors' are aware of, has measured down to these temperatures leaving it an open question as to whether \EGO\ exhibits long-range magnetic order - and at what temperature. Such information is important in studying magnetic frustration in the \textit{RE}PG and how it is tuned as the \textit{RE} moves across the Lanthanide period. 

To build on the results of Ghosh \textit{et al}., (Ref.~\onlinecite{Ghosh1998}) we performed susceptibility measurements on a powder sample of \EGO\ (Fig.~\ref{fig:two}). We observe no evidence of magnetic ordering down to 2 K. Above $\sim$ 100 K typical Curie-Weiss behavior is seen as reported for both \HGO\ and \DGO . Fitting the inverse susceptibility by the Curie-Wiess law ($\chi ^{-1} = C(T-\theta_{w})$) we obtain an effective magnetic moment of $\mu_{eff} = 9.8(3) \mu_B /\text{Er}$ and a Weiss temperature of $\theta_{w} = -14.4(2) \text{K}$. We find a negative Weiss temperature indicating antiferromagnetic (AFM) order in agreement with Ghosh \textit{et al.} \cite{Ghosh1998}. However, the absolute value of our Weiss temperature is larger than previously reported, we attribute this to the finer temperature steps and higher temperature cutoff (120 K) used in our analysis the former of which allows for the range of linear behavior to be more carefully determined \cite{Ghosh1998}.



\begin{figure}
	\includegraphics[width=\columnwidth]{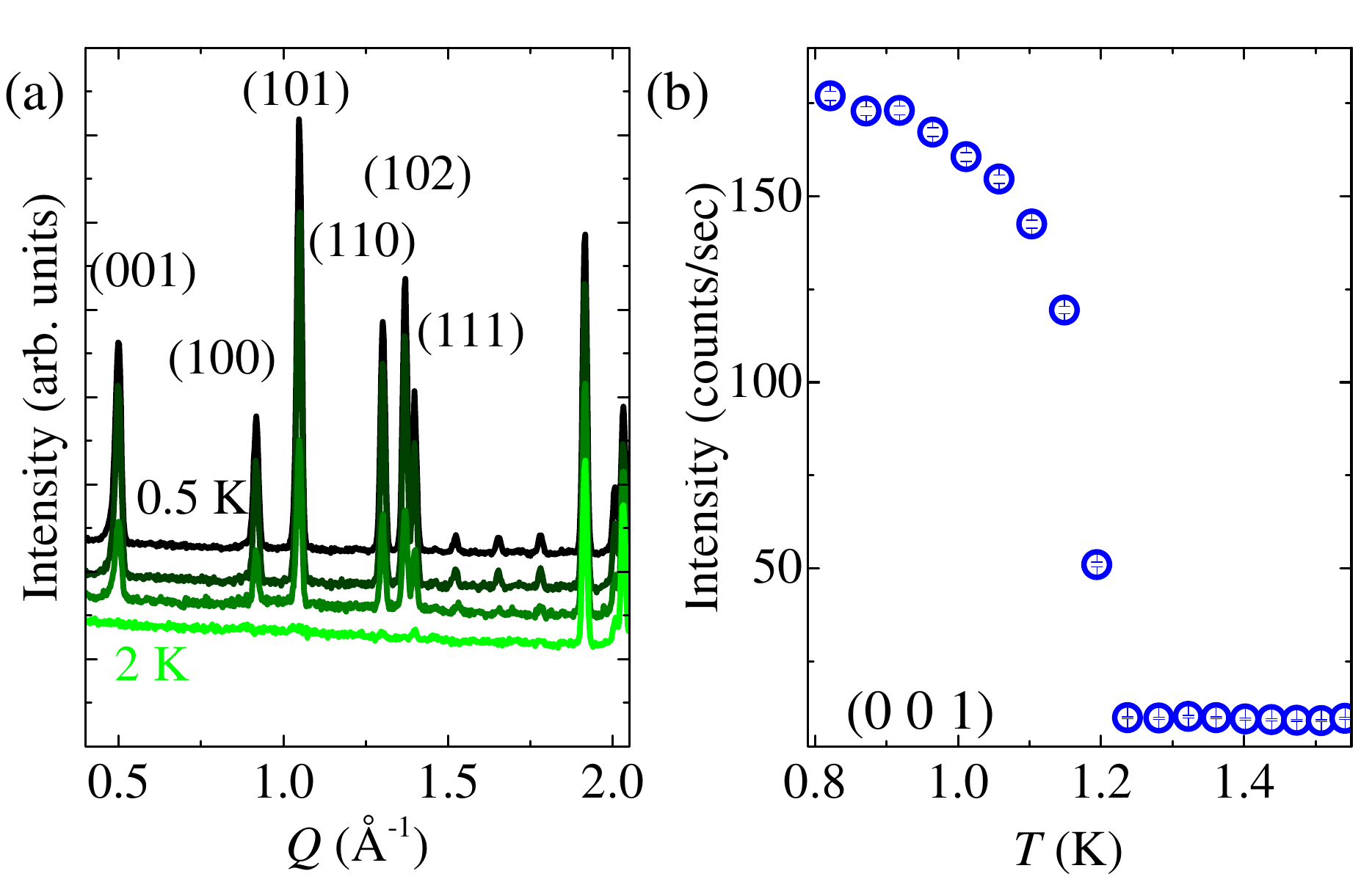}
	\caption{\label{fig:three} (a) Waterfall of neutron diffraction patterns collected at 2, 1.2, 0.8 and 0.5 K. (b) Temperature dependence of new low temperature (001) reflection collected on single crystal sample.}	
\end{figure}

In Figure~\ref{fig:one}(a), we noted an uncharacteristically large low-$Q$ background signal which monotonically gains intensity with decreasing $Q$. Such a signal is similar to that expected from the magnetic form factor of local moments in a paramagnetic state - in this case the tripositive Er$^{3+}$ $4f$ electrons  \cite{Moon1982}. At 2 K this signal is still present with a slightly increased count rate. As the material is cooled below 1.2 K a series of new low $Q$ peaks appear and the previously high background drops (Fig.~\ref{fig:three}(a)). Upon further cooling, the intensity of the new peaks grows until saturating at $\sim 0.8$ K, behavior indicative of magnetic ordering. Figure~\ref{fig:three}(b) shows the temperature dependence of the 0.5 \iA\ peak which may be fit with a power law ($M(T) \alpha (T - T_N)^{\beta}$) revealing the transition temperature to be $\sim$ 1.15 K. 

\begin{table}[h]
	\caption{\label{tab:two}Irreducible representations ($\Gamma$), magnetic space groups, magnetic supercell, number of basis vectors ($\psi$) and fit parameters for Rietveld refinements using 0.5 K data for magnetic orderings with $k = (000)$. The the low symmetry $P2_1^{'}$ model was not considered due to its significant increase in the number of refinable parameters.}
	\begin{ruledtabular}
		\begin{tabular}{lcccc}
     		 \multicolumn{1}{c}{$\Gamma$} & \multicolumn{1}{c}{Magnetic space group} & \multicolumn{1}{c}{$\psi$} & \multicolumn{1}{c}{$R_{wp}$} & \multicolumn{1}{c}{$\chi ^{2}$} \\
	\hline
	$\Gamma_1\ (mGM1)$ & $P4_12_12$ &  3 & 48.6 & 76.8 \\
	$\Gamma_2\ (mGM2)$ & $P4_1^{'}2_12^{'}$ & 3 & 9.76 & 3.10 \\
	$\Gamma_3\ (mGM3$) & $P4_12_1^{'}2^{'}$ & 3 & 49.8 & 80.7   \\
	$\Gamma_4\ (mGM4)$ & $P4_12_1^{'}2$ & 3 & 52.3 & 88.9   \\
	\multirow[t]{3}{*}{$\Gamma_5\ (mGM5)$} & $P2_1^{'}2_1^{'}2$ & 6 & 39.6 & 51.0 \\
	                        & $C22^{'}2_1^{'}$ & 6 & 40.4 & 53.3 \\
	                        & $P2_1^{'}$ & 12 &  & \\
		\end{tabular}
	\end{ruledtabular}
\end{table}

The drop in paramagnetic background at the ordering temperature and temperature dependence of the peak intensity are consistent with a magnetic origin to the signal. We therefore attempt to account for the new peaks with a magnetic model using represenational analysis, starting with the identification of an ordering vector. In this case, the numerous new low $Q$ peaks can be indexed with the nuclear unit cell indicating an ordering vector of $k = (0,0,0)$ (Fig.~\ref{fig:three}(a)). Using the SARAh software, the irreducible representations ($\Gamma$) consistent with the \Pftt\ space group symmetry and a $k = (0,0,0)$ ordering vector were generated and are shown in Table~\ref{tab:two} with the $\Gamma$ in Miller and Love notation \cite{Campbell2006} (for a complete description of the $\Gamma$ and constituent $\psi$ see the Supplemental Materials (SM))\cite{SM}. Five $\Gamma$ were found giving rise to seven possible magnetic structures. $\Gamma_1, \Gamma_2, \Gamma_3 \text{ and } \Gamma_4 $ each have three independent basis vectors ($\psi$) leading to three refinable parameters per structure - two for the \textit{ab} plane and one for the moment along the \textit{c} lattice direction. $\Gamma_5$ has twelve $\psi$ which can be subdivided by symmetry leading to three possible structures in the $\Gamma_5$ representation one of which is the low symmetry linear combination of all twelve $\psi$. Due to the exceptional increase in refinable parameters of this model we remove it from our considerations.   

\begin{figure}
	\includegraphics[width=\columnwidth]{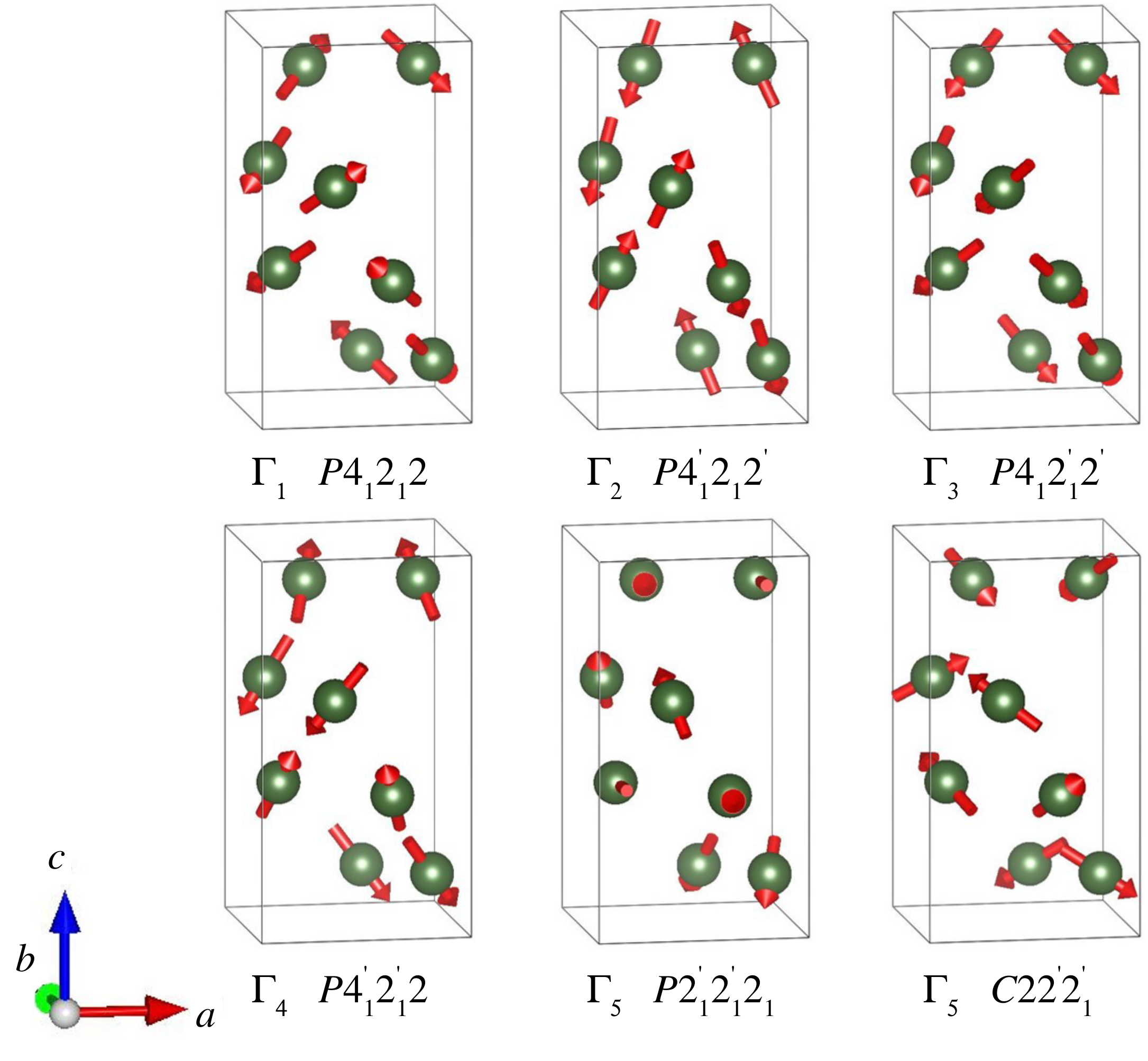}
	\caption{\label{fig:four} Possible magnetic structures with $k = (0,0,0)$ enumerated in Table~\ref{tab:two}. The values of the basis vectors where determined from attempted fitting of the 0.5 K neutron powder diffraction data.}	
\end{figure}

Figure~\ref{fig:four} shows the six remaining magnetic structures enumerated in Table~\ref{tab:two}. In earlier work on \HGO , Morosan \textit{et. al.} reported the $mGM1$ representation as accurately modeling the 1.36 K magnetic structure \cite{Morosan2008}. They determined an in-plane spiral structure with a $90$\degrees\ rotation between \textit{RE} planes. A version of this structure with an out-of-plane component is shown in the first panel of Fig.~\ref{fig:four}. 

In Table~\ref{tab:two} we report the fit residuals from Rietveld refinements using the 0.5 K data for each of the magnetic structures. The strong $(100)$ and $(110)$ peaks in Fig.~\ref{fig:three} indicate the presence of significant out-of-plane magnetic moment in \EGO .  Consequently, the in-plane $\Gamma_1$ model is incapable of adequately modeling our diffraction data. Even when an out-of-plane component is added,  $\Gamma_1$ results in a $R_{wp} > 40$ (we note that a fit to the 0.5 K data with no magnetic structure results in $R_{wp}$ and $\chi^{2}$ of 64.7 and 136 respectively). Similarly, the $\Gamma_3, \Gamma_4$ and $\Gamma_5$ models result in poor fits, with $R_{wp}$ never reaching $< 30$. Rather convincingly, we find the $\Gamma_2$ structure with magnetic space group symmetry $P4_{1}^{'}2_{1}2^{'}$ to produce the best fit parameters  with $R_{wp} = 9.76$ and $\chi^{2} = 3.10$. 

\begin{figure}
	\includegraphics[width=\columnwidth]{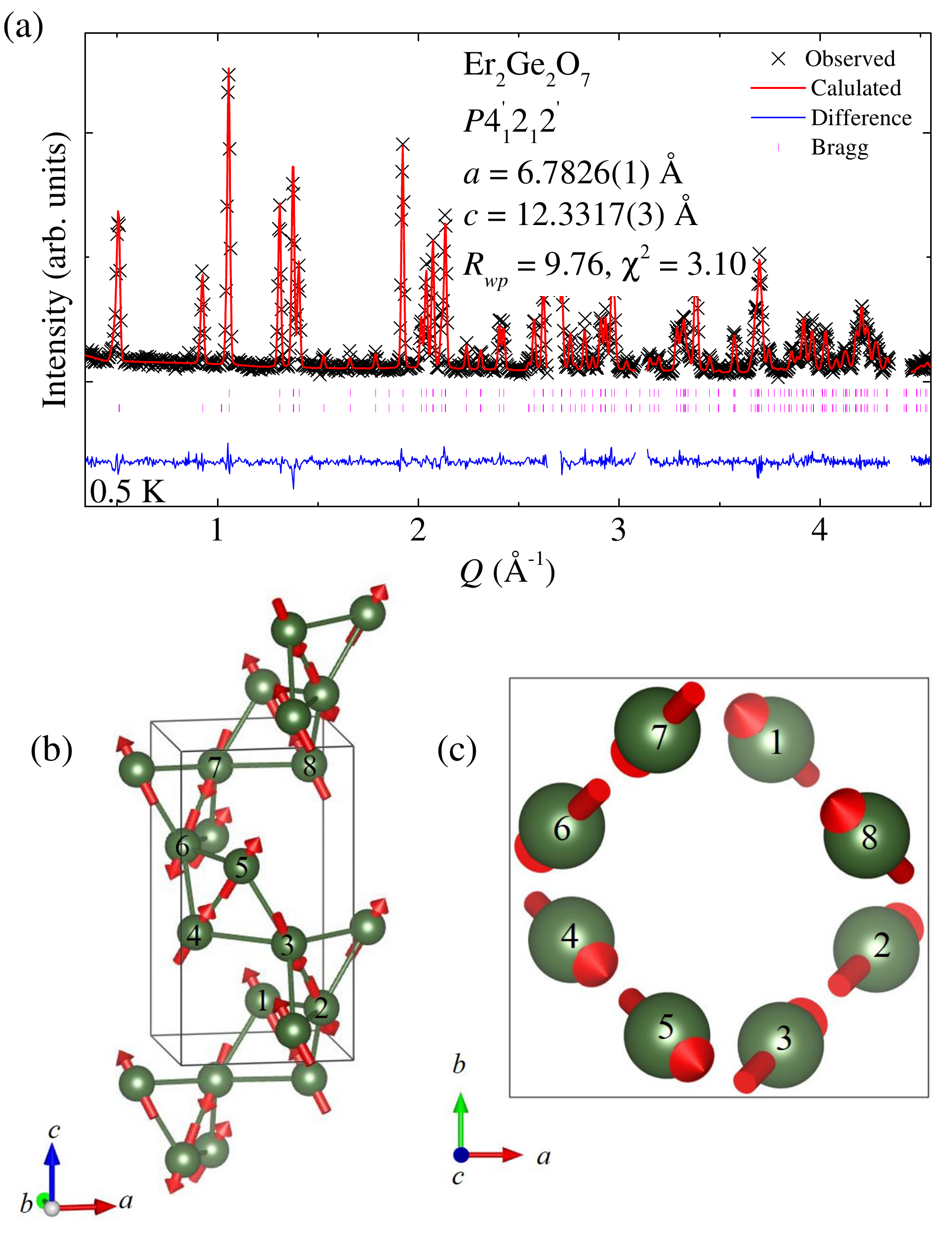}
	\caption{\label{fig:five} (a) Rietveld refinement using 0.5 K data and the magnetic structure generated by $\Gamma_3 $. Refined magnetic structure with Er atoms labeled as described in Fig.~\ref{fig:one} viewed along the (b) \textit{ab} plane and along the (c) \textit{c}-axis.}	
\end{figure}

The resulting fit is shown in Figure~\ref{fig:five}(a) producing a visually excellent agreement with the data. The fit \Gtw\ structure requires non-zero contributions from all three $\psi$ (two in-plane and one along the \c -axis) and leads to the three dimensional magnetic structure seen in Fig.~\ref{fig:five}(b) and (c). Considering the previously discussed Er edge-sharing sublattice, we find the magnetic structure has FM correlations along the edge-sharing Er-Er bond with the moment pointing along the bond direction. As the edge-sharing triangles are rotated to construct the full Er sublattice we find AFM correlations along \c\ between neighboring units and alternating FM and AFM correlations along the \a\ and \b\ directions creating a right handed spiral (Fig.~\ref{fig:five}(b)) with no net magnetization.

In this structure the Er moments exhibit a kind of \lq local Ising\rq\ behavior, where the moments are either spin-up or spin-down along the shortest and edge-sharing bond in the triangular Er sublattice. Notably, this is similar to the magnetic order of the spin-ice pyrochlores, where a large local anisotropy forces the \textit{RE} moments to point along the local $<111>$ direction either into or out of the \textit{RE} tetrahedra. Interestingly, it is this configuration which leads to the spin-ice rules in the prochlores. As will be discussed more later, the observation of a similar \lq local Ising\rq\ behavior indicates similar physics in \EGO . 

Our refinements produce a magnetic moment of $8.1(3) \mu_B/Er$ with $m_a = 2.66(4)$, $m_b = 3.03(3)$ and $m_c = -6.98(4) \mu_b/$Er. For elemental tripositive Er (Er$^{3+}$) which has total angular momentum $J$ of $\frac{15}{2}$, the theoretical saturated magnetic moment is $9 \mu_B/$Er - close to our refined value. We note that unlike \HGO , the moment size predicted from magnetization measurements agrees reasonably well with our results. 

Using the $\theta_{w} = -14.4$ K together with \Tn\ = 1.15 K we can determine the frustration index $f = |\theta_{W}|/T_N \sim 13$ indicating significant frustration \cite{Ramirez1994,Balents2010}. This is higher than the $f \sim\ 6 \text{ and } 2$ found for \HGO\ and \DGO\ respectively \cite{Morosan2008, Ke2008}. With a motif of edge sharing triangles, the Er sublattice might be expected to show frustrated behavior, however, as was the case in \HGO , the frustration is apparently alleviated by the formation of a spiral magnetic structure. 

Despite being isostructural, we find that the magnetic behavior in the \textit{RE}PG as the \textit{RE} is incremented from Dy to Ho to Er somewhat complex. For all three, the reported magnetic moment has been near the saturated free ion expectation ($\sim $10.5, 9.1 and 8.1 $\mu_B$ respectively) \cite{Morosan2008,Ke2008}. Furthermore, the magnetic ordering temperature follows a similar trend, decreasing with decreasing moment size from 2.2 to 1.6 to 1.15 K. Although the crystallographic parameters of the Dy compound are not reported, the decrease in \Tn\ correlates with the lattice contraction anticipated for the reduction in ionic radius from Dy to Ho to Er \cite{Shannon1976}. Interestingly, the $\theta_{w}$ follows the opposite trend reaching a maximum absolute value of 14.4 K for the Er compound and decreasing to 9.6 and 4.4 K for Ho and Dy respectively \cite{Morosan2008,Ke2008,Ghosh1998}. Considering the complex dynamic magnetic behavior reported in both \HGO\ and \DGO\ which indicated multiple competing interactions and time-scales, it is unsurprising that simple moment size considerations are inadequate here for predicting the bulk magnetic behavior which is likely tuned by subtle effects in the ErO$_7$ local environment and the corresponding crystal field levels.  

\subsection{\label{subsec:mag} Field Dependence of the Magnetic Structure}

In light of the strong field dependent behavior reported for both \HGO\ and \DGO , neutron diffraction patterns of \EGO\ were collected under an applied field \cite{Morosan2008, Ke2008}. In a powder sample, such study is useful as a first approach to determine field dependence and critical values, though the random orientation of the crystallites in the powder suggests that the measured state is likely a mixture of different states. 

\begin{figure}
	\includegraphics[width=\columnwidth]{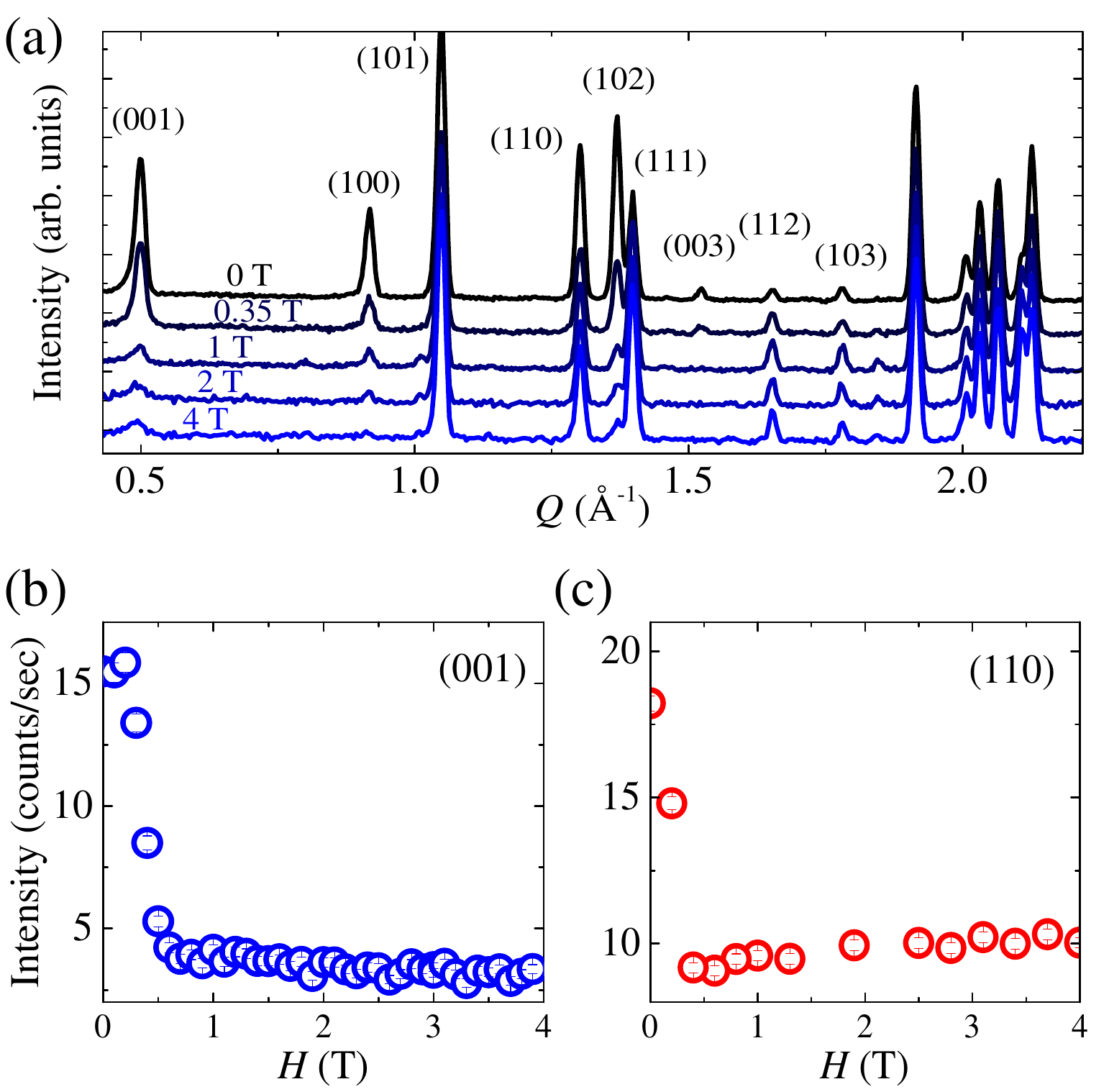}
	\caption{\label{fig:six} (a) Waterfall of neutron diffraction patterns collected at 0.5 K under 0, 0.35, 1, 2 and 4 T applied fields. Field dependence of the (b) (001) and (c) (110) peak intensities. All panels are of diffraction data collected using powder samples.}	
\end{figure}

Figure~\ref{fig:six}(a) shows powder diffraction patterns collected at 0.5 K under fields of 0, 0.35, 1, 2 and 4 T. At the lowest applied field of 0.35 T, changes in the intensities of the magnetic peaks are observed, with a series of reflections (e.g. (001), (100) and (102)) losing intensity. As the field is increased to 4 T, these reflections continue to monotonically decrease in intensity. At 4 T the (001), (100) and (102) reflections are almost entirely suppressed with integrated intensities less than 10\%\ of their zero-field value.  Simultaneously, the (111), (112) and (103) reflections gain intensity for $H > 0$ T. These reflections increase monotonically as the applied field is increased to 4 T (Fig.~\ref{fig:six}(a)).

The increasing/decreasing intensity of different species of reflections indicates the applied field is driving a magnetic transition and not suppressing the magnetic order. However as seen in Fig.~\ref{fig:six}(a), no new reflections arise under field. Therefore, the new structure must share the original $k = (0,0,0)$ ordering vector. Fig.~\ref{fig:six}(b) and (c) show order parameter scans for the (001) and (110) reflections performed as a function of applied field. For the (001) reflection, the peak intensity steeply drops showing critical behavior with a transition field of $ H_{c} \sim\ 0.5$ T. Similarly albeit on a coarser field grid, the (110) reflection is quickly suppressed and nearly constant by 0.5 T again showing critical behavior. These observations suggest a true phase transition. Considering the $k$ vector, we should expect the new magnetic structure to select a different irrep or combination of irreps from Table~\ref{tab:two} (Fig.~\ref{fig:four}). 

Such a field driven magnetic transition can be characterized as a metamagnetic transition (MMT) and should result from a  reorientation of the spin directions in response to the applied field and the anisotropies of the magnetic site \cite{Stryjewski1977}. This could result in either a spin-flip transition, where the moments undergo a 180\degrees\ change in polarization along the direction of the applied field, or a spin-flop transition where the moments rotate in response to the field. The former outcome is expected for a material with strong anisotropies while the latter is suggestive of weaker anisotropies \cite{Stryjewski1977}. Therefore, characterization of the resulting magnetic structure would give information about the anisotropy of the rare-earth site. However, this is not possible with powder diffraction data for reasons described previously and so motivates single crystal neutron diffraction study.  

We note that the behavior of the order parameters shown in Fig.~\ref{fig:six}(b) and (c) above $H_{c}$ is perhaps not quite constant as one would expect. We attribute this to the lack of orientation to the applied field due to the polycrystalline nature of the powder sample. A field driven transition in a system with high anisotropy should be dependent on the direction of the applied field \cite{Fukazawa2002}. Therefore it is possible, we are averaging over different field induced magnetic structures, or seeing other effects due to a \lq misaligned\rq\ field \cite{Fukazawa2002}.

\begin{table}
	\caption{\label{tab:three}Irreducible representations, magnetic space groups and fit parameters for single crystal Rietveld refinements for data collected at 0.5 K under applied fields of 1 and 4 T. The irrep labels are consistent with those presented in Table~\ref{tab:two}.}
	\begin{ruledtabular}
		\begin{tabular}{lccc}
     		 \multicolumn{1}{c}{$\Gamma$} & \multicolumn{1}{c}{Magnetic space group} & \multicolumn{2}{c}{$R_{wF^{2}}$($\chi ^{2}$)} \\
     	   	\multicolumn{2}{c}{} & \multicolumn{1}{c}{ 1 T} & \multicolumn{1}{c}{4 T} \\
	\hline
	
	$\Gamma_1\ (mGM1)$ & $P4_12_12$  & 79.0(11.5) & 99.8(16.0) \\
	$\Gamma_2\ (mGM2)$ & $P4_1^{'}2_12^{'}$ & 45.9(3.87) & 87.5(14.0) \\
	$\Gamma_3\ (mGM3$) & $P4_12_1^{'}2^{'}$ & 42.5(3.31) & 77.4(11.0)  \\
	$\Gamma_4\ (mGM4)$ & $P4_12_1^{'}2$ & 36.0(2.38) & 61.2(6.87)   \\
	\multirow[t]{2}{*}{$\Gamma_5\ (mGM5)$} & $P2_1^{'}2_1^{'}2$ & 27.9(2.30) & 13.9(0.62) \\
	                        & $C22^{'}2_1^{'}$ & 31.2(2.04) & 44.9(3.69) \\
		\end{tabular}
	\end{ruledtabular}
\end{table}

Single crystal neutron diffraction was performed to elucidate the field dependent magnetic structure. Geometrical considerations of the experimental setup limited access of reciprocal space to a single ($HKL$) plane perpendicular to the applied field. The choices of scattering plane and field direction where therefore limited. The previous work on \DGO\ and \HGO\ identified the easy axis as in the \textit{ab} plane \cite{Ke2008, Morosan2008}. Furthermore, our neutron powder diffraction data demonstrated ($H00$)/($0K0)$) and ($00L$) type peaks as responding strongly to field. Therefore, we chose to align our applied field with the crystallographic \textit{a} direction and probe a ($H0L$) cut of reciprocal space \cite{Ke2008, Morosan2008}.  

\begin{figure}
	\includegraphics[width=\columnwidth]{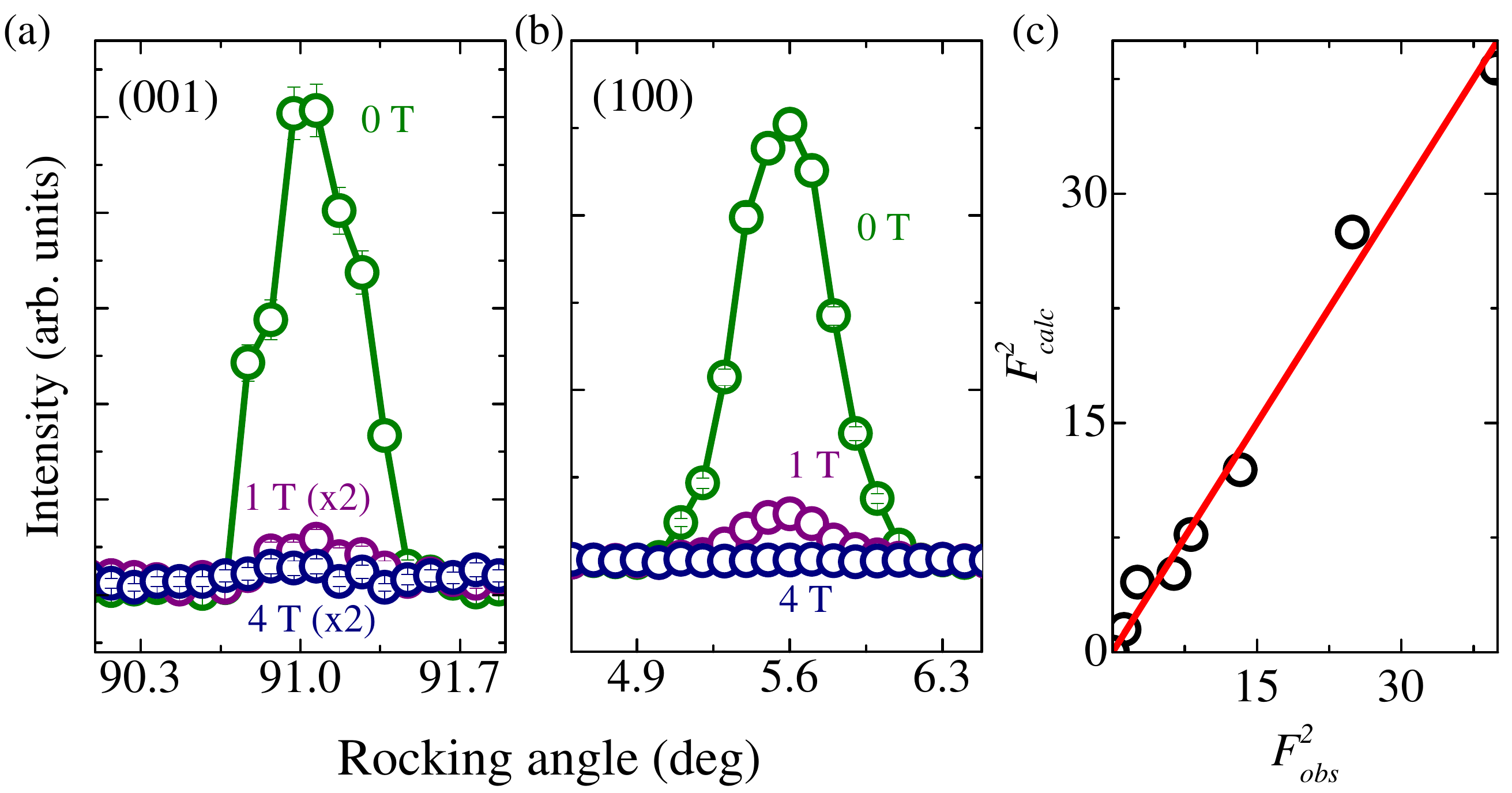}
	\caption{\label{fig:seven} Single crystal data showing rocking curves on the (a) (001) and (b) (100) magnetic reflections under applied fields of 0, 1 and 4 T. (c) Calculated peak intensities ($F^{2}_{calc}$) plotted as a function of the observed intensities ($F^{2}_{obs}$) for Rietveld refinements using the $P2_{1}^{'}2_{1}^{'}2$ magnetic structure to model the single crystal data collected under a 4 T applied field.}	
\end{figure}

Fig.~\ref{fig:seven}(a) and (b) show the field dependence of the (001) and (100) magnetic reflections respectively. In agreement with our powder results, both peaks are suppressed by the applied field with the largest intensity changes between 0 and 1 T. As for the powder experiment, with increasing field both reflections are monotonically suppressed. In the single crystal however, the intensities of these peaks are more strongly suppressed and by 4 T both are background equivalent. 

In order to perform Rietveld refinements, rocking curves were collected on $\sim\ 25$ nuclear and magnetic reflections under applied fields of 0, 1, and 4 T. The integrated intensities were then used to model the possible magnetic structures  allowed by the $k=(000)$ ordering vector (Fig.~\ref{fig:four}). Due to the relative scarcity of peaks, the nuclear structure was fixed at the crystallographic properties determined from the low temperature zero field powder measurements. As a check, we modeled the zero field structure and confirmed the results of our powder studies finding the $\Gamma _{2}$ irrep as producing the best fit ($R_{wF^{2}} \sim\ 14\% $). 

The resulting fit residuals for fits to the 1 and 4 T data are shown in Table~\ref{tab:four}. Starting with the 4 T data, we find the $\Gamma _{2}$ model is no longer able to reproduce the measured intensities indicating, as suggested by the order parameter scans, the sample has undergone a phase transition. Comparing the fit residuals for the various models we find only magnetic structures within the $\Gamma _{5}$ irrep produce reasonable fits. Of those the magnetic space group $P2_{1}^{'}2_{1}^{'}2$ produces the lowest fit residuals with similar quality to our known zero field structure (the calculated and observed intensities are shown in Fig~\ref{fig:seven}(c)). We note that because of the limited number of peaks we did not attempt modeling with the $\Gamma _{5}$ $P2_{2}^{'}$ structure for the same reasons previously discussed for the powder modeling. 

Considering the 1 T data, we obtain similar results albeit with larger residuals. As seen in Fig.~\ref{fig:six}(d) and (c), at 1 T there is still intensity on the (100) and (001) reflections which is gone by 4 T. This may indicate that the phase transition is not complete at this intermediate field, possibly due to small mis-alignments of our co-aligned crystals which results in some remnant of the sample remaining in the zero field magnetic structure. Nonetheless, the  $P2_{1}^{'}2_{1}^{'}2$ structure produces the lowest fit residuals for both the 1 and 4 T data.  

\begin{figure}
	\includegraphics[width=\columnwidth]{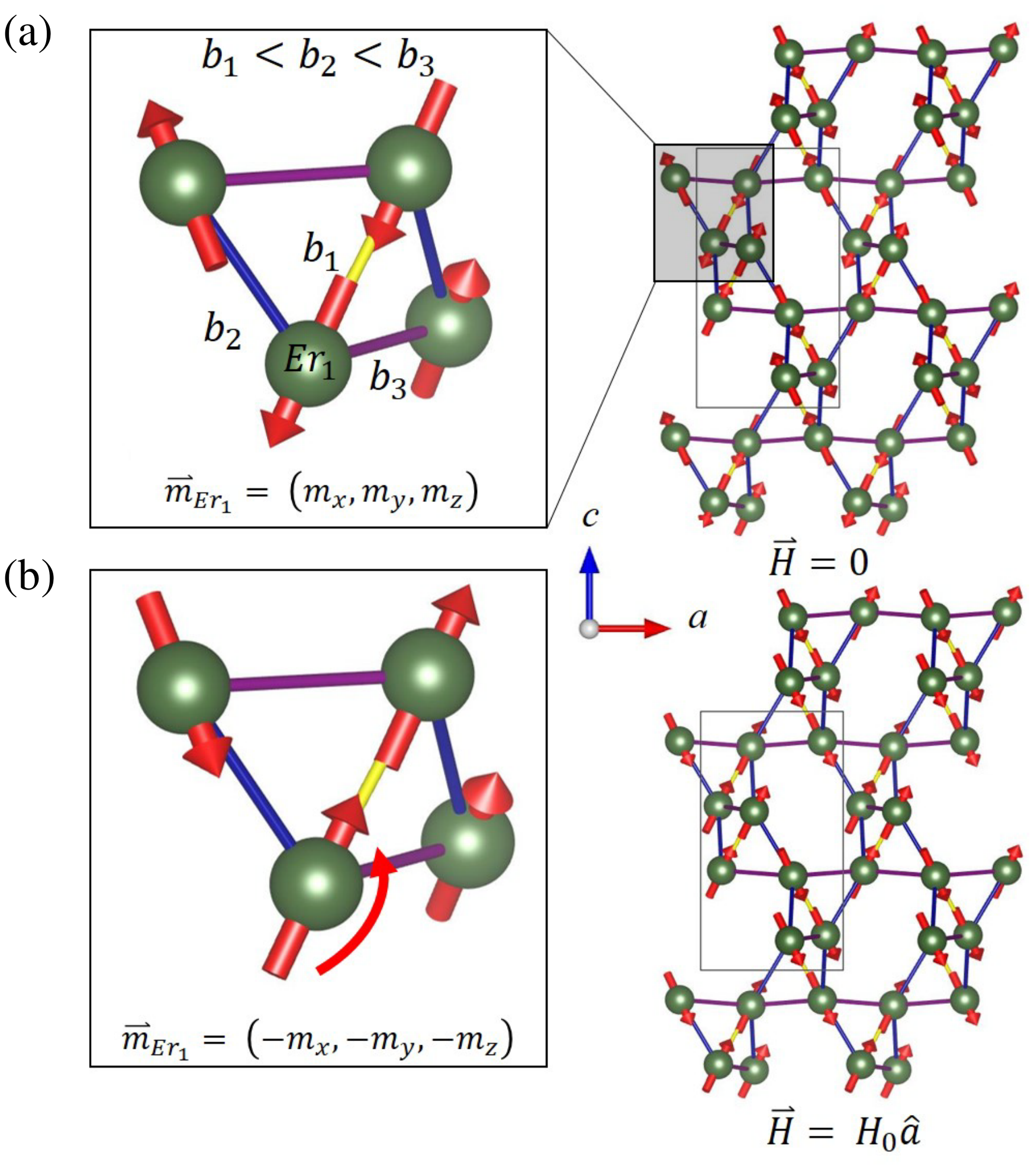}
	\caption{\label{fig:eight} Best fit magnetic structure for the (a) 0 and (b) 4 T data (with field applied along the \a -axis. for each model a zoomed in region of the Er sublattice is shown emphasizing the geometry of the edge sharing Er triangles. The labels $b_1$, $b_2$ and $b_3$ refer to the three different Er-Er distances within each pair of edge sharing Er triangles. The label Er$_1$ only refers to this figure for the purpose of describing the inversion of the magnetic moment. The $b_1$ short Er-Er distance is denoted by a yellow bar while the longer $b_2$ and $b_3$ distances are denoted by blue and purple bars respectively. This allows for the short Er-Er distance to be seen easily throughout the structure models.}	
\end{figure}

Figure~\ref{fig:eight} shows the both the 0 and 4 T magnetic structures determined using the single crystal data. Comparing these reveals an interesting similarity - they are nearly identical save for an inversion of moments with components anti parallel to the applied field about their crystallographic position. The left panels of Fig.~\ref{fig:eight} show this more clearly by focusing on a single pair of edge sharing Er$_4$ triangles. Each Er$_4$ unit has three distinct bond lengths labeled $b_1 = 3.57(1) $, $b_2 = 3.60(1)$ and $b_3 = 3.80(1)$, with the latter two defining the sides of the unit and $b_1$ the shared edge internal to the Er$_4$ unit. As previously discussed, along the short $b_1$ the Er moments' are coaxial with the bond, this is seen in both the 0 and 4 T structures. Considering the full unit cell, this Er$_4$ unit forms the Er sublattice by connecting via a corner along the long axis to another Er$_4$ unit's short axis (\textit{i.e.} $b_1$). In this way every, Er is part of the short axis of an Er$_4$ unit and has its moment subject to the \rq local Ising\lq\ behavior - always aligning with $b_1$. 

In the applied field, the Er$_4$ may be divided into two sets - one with the \textit{a} component of the $b_1$ Er aligned with the field, the other with the \textit{a} component of the $b_1$ Er anti-parallel. In the later, there exists competition between the energetically unfavorable field alignment and the strong local anisotropy fixing the moment to the $b_1$ direction. As the field is increased the anti-parallel moments invert reducing the misalignment with the applied field while not breaking from the easy axis. Remarkably, Er moments which have no component anti-parallel to the field appear unaffected within the statistical certainty of our measurement.  
     
It is significant that we observe complete inversion of the anti-parallel moments as it indicates a strong anisotropy which keeps the moment locked to a certian axis. Such a response to the applied field describes a class I MMT \cite{Stryjewski1977}. Class I MMTs, or spin-flip transitions, are characterized by a strong local anisotropy on the magnetic site which prevents the moment from rotating away from the easy axis (i.e. spin-flop) in response to the applied field. In class I MMT's, the anisotropy requires the moment to be either parallel or anitparallel to the easy-axis. Consequently, once a critical field is reached ($H_c$) the material undergoes a first-order phase transition to a ferrimagnetic or paramagnetic state (the details of the $HvT$ phase diagram are variable but at low temperatures the MMT is discontinuous) \cite{Stryjewski1977}. This is exactly what is shown in our analysis. 

To corroborate this classification we can attempt to characterize the order of the observed MMT transition. While the most straightforward route would be to consider the peak intensities as the magnetic order parameter ($Intensity \propto M^2$) unfortunately, the order parameter scans shown in Fig.~\ref{fig:six}(b) and (c) are not reliable in this regard due to low data-point density and powder origin. However, symmetry considerations of the magnetic structures are helpful. The change from the zero-field ($\Gamma _{2}$) irrep to the in-field structure of $\Gamma _{5}$ is not a continuous transition in which case we would expect one irrep or an associated basis vector to develop a finite value. Rather what we see is a change in irreps, where one irrep cannot smoothly evolve into the other. From this we can determine that the transition is of first-order in agreement with the general theory of class I MMTs. 

We note that the manifestation of the MMT in \EGO\ is interesting - in most metamagnets the magnetic moments exhibit a single universal easy axis leading to co-linear or co-planar magnetic structures. This is the case of the zero field structure of \HGO\ and presumably \DGO . However, in \EGO\ the moments have a spiral structure with significant out-of-plane canting. Explicating the rare-earth magnetic Hamiltonian would be useful to understanding this change in magnetic structure but is outside the scope of the present work, however we posit that it may be due to the change in \textit{f}-orbital shape from oblate in Ho and Dy to prolate in Er. Nonetheless, such canting is a special case of MMT \cite{Kitano1964}. 

Furthermore, we notice a similarity to the spin-ice materials in the presence of a MMT with a relatively low critical field of $H_c < 1$ T (compared to $H_c >> 1$ T seen in many metamagnetic materials)\cite{Deppe2012,Medrano2017,Stryjewski1977,Harris1997,Sibille2016}.  In Ho$_2$Ti$_2$O$_7$ a similar $H_c \sim 0.3$T and strong local anisotropy were observed . In this canonical spin-ice, the MMT revealed the competition between FM exchange and single-ion anisotropy with the low-field overcoming the exchange interaction - giving important information about the underlying physics.  More broadly, in the spin-ice pyrochlores, the MMT from an unordered zero-field state to a magnetically ordered state in applied fields is intimately related to the novel physics being a magnifestation of a liquid-gas transition of emergent magnetic monopoles \cite{Castelnovo2008}. While the behavior we report here is different, having a fully ordered AFM ground state, the similarities in MMT indicate similar competing interactions. 

Indeed, in further analogy here we find also the easy axis is determined locally \cite{Fukazawa2002, Harris1997, Castelnovo2008}. This is a vital component in the pyrochlores to the establishment of the frustrating ice rules. Interestingly, the \textit{RE}PG structure is ostensibly tunable - related to the more usual planar spiral structure exhibited in both \HGO\ and \DGO\ only by changing the \textit{RE}. Whether this is an effect of the smaller ionic size - which as discussed previously, results in shorter \textit{RE}-\textit{RE} distances and therefore presumably stronger magnetic dipole interactions - of affected bond angles and exchange interactions, changes in the magnetic properties of the \textit{RE} ion due to different spin-quantum numbers, resultant changes to the crystal field levels or combinations of all of these effects is unknown and warrants continued study. Determining how to tune these parameters to increase the frustration in these \textit{RE}PG is of great interest to possibly find a new system with spin-ice or spin-liquid physics - especially in light of the previous work which has shown the spins to be Ising-like and dynamic susceptibilities reminiscent of spin-freezing physics \cite{Morosan2008, Ke2008}. 

\section{\label{sec:con} Conclusions}

We have reported on neutron diffraction studies elucidating the zero-field and field dependent low temperature magnetic structures of \textit{RE}PG \EGO . In zero-field we find magnetic ordering at $T_N = 1.15$ K which together with a $\theta _w = -14.4$ K indicates a relatively strong frustration index of $f \sim\ 13$. All magnetic reflections can be indexed with a $k = (0,0,0)$ AFM structure and modeled well by magnetic space group symmetry $P4_1^{'}2_12^{'}$ with Er moments of $m = 8.1 \mu_B$ aligned along the short Er-Er distance of the \textit{RE} sublattice. Field dependent studies reveal a metamagnetic transition with a critical field of $H_c \sim 0.35$ T at 0.5 K which is stable up to applied fields of 4 T. This second magnetic structure maintains $k = (0,0,0)$ but with a spin reorientation to magnetic space group symmetry $P2_{1}^{'}2_{1}^{'}2$ driven by an inversion of Er sites with magnetic moment components antiparallel to the applied field. Symmetry analysis of these two structures indicate the field driven transition as a first-order class I spin-flip, metamagnetic transition characteristic of a strong local anisotropy at the magnetic site. Together the observations of a local easy axis along the short Er-Er distance and a spin-flip transition encourage a \lq local Ising\rq\ type description of the magnetic order in which each Er moment aligns either spin-up or spin-down along this local easy-axis.      

We opine that the observation of both a metamagnetic transition and local Ising behavior is interesting and indicates physics analogous to the spin-ice pyrochlores in considerably different \textit{RE} sublattice. In many of the spin-ice pyrochlores such metamagnetic transitions have been observed but from paramagnetic to ferromagnetic rather than from antiferromagnetic as seen here. In those systems, such a transition is interpreted as the tunable gas-liquid transition of magnetic monopoles. While such a situation is not the case in \EGO\, the spin-flip transition is evidence of strong anisotropy and quantum behavior as the spins are locked along discrete directions and can be tuned selectively between spin-up and spin-down states with an applied field.  This observation encourages further work looking to explicate the spin Hamiltonian. Such understanding may lead to new ways to tune the magnetic structure - either gaining finer control over the individual spin states or possibly increasing the frustration and achieving a spin-ice state. Such work is of potential interest in the search for new materials with exotic quantum ground states such as spin-liquids.

\begin{acknowledgments}
The part of the research that was conducted at ORNL’s High Flux Isotope Reactor was sponsored by the Scientific User Facilities Division, Office of Basic Energy Sciences, US Department of Energy. The research is partly supported by the U.S. Department of Energy (DOE), Office of Science, Basic Energy Sciences (BES), Materials Science and Engineering Division. Work performed at Clemson University was funded by DOE BES Grant DE-SC0014271.
\end{acknowledgments}


\begin{thebibliography}{49}%
\makeatletter
\providecommand \@ifxundefined [1]{%
 \@ifx{#1\undefined}
}%
\providecommand \@ifnum [1]{%
 \ifnum #1\expandafter \@firstoftwo
 \else \expandafter \@secondoftwo
 \fi
}%
\providecommand \@ifx [1]{%
 \ifx #1\expandafter \@firstoftwo
 \else \expandafter \@secondoftwo
 \fi
}%
\providecommand \natexlab [1]{#1}%
\providecommand \enquote  [1]{``#1''}%
\providecommand \bibnamefont  [1]{#1}%
\providecommand \bibfnamefont [1]{#1}%
\providecommand \citenamefont [1]{#1}%
\providecommand \href@noop [0]{\@secondoftwo}%
\providecommand \href [0]{\begingroup \@sanitize@url \@href}%
\providecommand \@href[1]{\@@startlink{#1}\@@href}%
\providecommand \@@href[1]{\endgroup#1\@@endlink}%
\providecommand \@sanitize@url [0]{\catcode `\\12\catcode `\$12\catcode
  `\&12\catcode `\#12\catcode `\^12\catcode `\_12\catcode `\%12\relax}%
\providecommand \@@startlink[1]{}%
\providecommand \@@endlink[0]{}%
\providecommand \url  [0]{\begingroup\@sanitize@url \@url }%
\providecommand \@url [1]{\endgroup\@href {#1}{\urlprefix }}%
\providecommand \urlprefix  [0]{URL }%
\providecommand \Eprint [0]{\href }%
\providecommand \doibase [0]{http://dx.doi.org/}%
\providecommand \selectlanguage [0]{\@gobble}%
\providecommand \bibinfo  [0]{\@secondoftwo}%
\providecommand \bibfield  [0]{\@secondoftwo}%
\providecommand \translation [1]{[#1]}%
\providecommand \BibitemOpen [0]{}%
\providecommand \bibitemStop [0]{}%
\providecommand \bibitemNoStop [0]{.\EOS\space}%
\providecommand \EOS [0]{\spacefactor3000\relax}%
\providecommand \BibitemShut  [1]{\csname bibitem#1\endcsname}%
\let\auto@bib@innerbib\@empty
\bibitem [{\citenamefont {Wannier}(1950)}]{Wannier1950}%
  \BibitemOpen
  \bibfield  {author} {\bibinfo {author} {\bibfnamefont {G.~H.}\ \bibnamefont
  {Wannier}},\ }\href {\doibase 10.1103/PhysRev.79.357} {\bibfield  {journal}
  {\bibinfo  {journal} {Phys. Rev.}\ }\textbf {\bibinfo {volume} {79}},\
  \bibinfo {pages} {357} (\bibinfo {year} {1950})}\BibitemShut {NoStop}%
\bibitem [{\citenamefont {Derrida}\ \emph {et~al.}(1978)\citenamefont
  {Derrida}, \citenamefont {Vannimenus},\ and\ \citenamefont
  {Pomeau}}]{Derrida1978}%
  \BibitemOpen
  \bibfield  {author} {\bibinfo {author} {\bibfnamefont {B.}~\bibnamefont
  {Derrida}}, \bibinfo {author} {\bibfnamefont {J.}~\bibnamefont {Vannimenus}},
  \ and\ \bibinfo {author} {\bibfnamefont {Y.}~\bibnamefont {Pomeau}},\ }\href
  {http://stacks.iop.org/0022-3719/11/i=23/a=019} {\bibfield  {journal}
  {\bibinfo  {journal} {Journal of Physics C: Solid State Physics}\ }\textbf
  {\bibinfo {volume} {11}},\ \bibinfo {pages} {4749} (\bibinfo {year}
  {1978})}\BibitemShut {NoStop}%
\bibitem [{\citenamefont {Ramirez}(1994)}]{Ramirez1994}%
  \BibitemOpen
  \bibfield  {author} {\bibinfo {author} {\bibfnamefont {A.}~\bibnamefont
  {Ramirez}},\ }\href {\doibase 10.1146/annurev.ms.24.080194.002321} {\bibfield
   {journal} {\bibinfo  {journal} {Annual Review of Materials Science}\
  }\textbf {\bibinfo {volume} {24}},\ \bibinfo {pages} {453} (\bibinfo {year}
  {1994})}\BibitemShut {NoStop}%
\bibitem [{\citenamefont {Gardner}\ \emph {et~al.}(2010)\citenamefont
  {Gardner}, \citenamefont {Gingras},\ and\ \citenamefont
  {Greedan}}]{Gardner2010}%
  \BibitemOpen
  \bibfield  {author} {\bibinfo {author} {\bibfnamefont {J.~S.}\ \bibnamefont
  {Gardner}}, \bibinfo {author} {\bibfnamefont {M.~J.~P.}\ \bibnamefont
  {Gingras}}, \ and\ \bibinfo {author} {\bibfnamefont {J.~E.}\ \bibnamefont
  {Greedan}},\ }\href {\doibase 10.1103/RevModPhys.82.53} {\bibfield  {journal}
  {\bibinfo  {journal} {Rev. Mod. Phys.}\ }\textbf {\bibinfo {volume} {82}},\
  \bibinfo {pages} {53} (\bibinfo {year} {2010})}\BibitemShut {NoStop}%
\bibitem [{\citenamefont {Harris}\ \emph {et~al.}(2018)\citenamefont {Harris},
  \citenamefont {Sato}, \citenamefont {Berkley}, \citenamefont {Reis},
  \citenamefont {Altomare}, \citenamefont {Amin}, \citenamefont {Boothby},
  \citenamefont {Bunyk}, \citenamefont {Deng}, \citenamefont {Enderud},
  \citenamefont {Huang}, \citenamefont {Hoskinson}, \citenamefont {Johnson},
  \citenamefont {Ladizinsky}, \citenamefont {Ladizinsky}, \citenamefont
  {Lanting}, \citenamefont {Li}, \citenamefont {Medina}, \citenamefont
  {Molavi}, \citenamefont {Neufeld}, \citenamefont {Oh}, \citenamefont
  {Pavlov}, \citenamefont {Perminov}, \citenamefont {Poulin-Lamarre},
  \citenamefont {Rich}, \citenamefont {Smirnov}, \citenamefont {Swenson},
  \citenamefont {Tsai}, \citenamefont {Volkmann}, \citenamefont {Whittaker},\
  and\ \citenamefont {Yao}}]{Harris2018}%
  \BibitemOpen
  \bibfield  {author} {\bibinfo {author} {\bibfnamefont {R.}~\bibnamefont
  {Harris}}, \bibinfo {author} {\bibfnamefont {Y.}~\bibnamefont {Sato}},
  \bibinfo {author} {\bibfnamefont {A.~J.}\ \bibnamefont {Berkley}}, \bibinfo
  {author} {\bibfnamefont {M.}~\bibnamefont {Reis}}, \bibinfo {author}
  {\bibfnamefont {F.}~\bibnamefont {Altomare}}, \bibinfo {author}
  {\bibfnamefont {M.~H.}\ \bibnamefont {Amin}}, \bibinfo {author}
  {\bibfnamefont {K.}~\bibnamefont {Boothby}}, \bibinfo {author} {\bibfnamefont
  {P.}~\bibnamefont {Bunyk}}, \bibinfo {author} {\bibfnamefont
  {C.}~\bibnamefont {Deng}}, \bibinfo {author} {\bibfnamefont {C.}~\bibnamefont
  {Enderud}}, \bibinfo {author} {\bibfnamefont {S.}~\bibnamefont {Huang}},
  \bibinfo {author} {\bibfnamefont {E.}~\bibnamefont {Hoskinson}}, \bibinfo
  {author} {\bibfnamefont {M.~W.}\ \bibnamefont {Johnson}}, \bibinfo {author}
  {\bibfnamefont {E.}~\bibnamefont {Ladizinsky}}, \bibinfo {author}
  {\bibfnamefont {N.}~\bibnamefont {Ladizinsky}}, \bibinfo {author}
  {\bibfnamefont {T.}~\bibnamefont {Lanting}}, \bibinfo {author} {\bibfnamefont
  {R.}~\bibnamefont {Li}}, \bibinfo {author} {\bibfnamefont {T.}~\bibnamefont
  {Medina}}, \bibinfo {author} {\bibfnamefont {R.}~\bibnamefont {Molavi}},
  \bibinfo {author} {\bibfnamefont {R.}~\bibnamefont {Neufeld}}, \bibinfo
  {author} {\bibfnamefont {T.}~\bibnamefont {Oh}}, \bibinfo {author}
  {\bibfnamefont {I.}~\bibnamefont {Pavlov}}, \bibinfo {author} {\bibfnamefont
  {I.}~\bibnamefont {Perminov}}, \bibinfo {author} {\bibfnamefont
  {G.}~\bibnamefont {Poulin-Lamarre}}, \bibinfo {author} {\bibfnamefont
  {C.}~\bibnamefont {Rich}}, \bibinfo {author} {\bibfnamefont {A.}~\bibnamefont
  {Smirnov}}, \bibinfo {author} {\bibfnamefont {L.}~\bibnamefont {Swenson}},
  \bibinfo {author} {\bibfnamefont {N.}~\bibnamefont {Tsai}}, \bibinfo {author}
  {\bibfnamefont {M.}~\bibnamefont {Volkmann}}, \bibinfo {author}
  {\bibfnamefont {J.}~\bibnamefont {Whittaker}}, \ and\ \bibinfo {author}
  {\bibfnamefont {J.}~\bibnamefont {Yao}},\ }\href {\doibase
  10.1126/science.aat2025} {\bibfield  {journal} {\bibinfo  {journal}
  {Science}\ }\textbf {\bibinfo {volume} {361}},\ \bibinfo {pages} {162}
  (\bibinfo {year} {2018})}\BibitemShut {NoStop}%
\bibitem [{\citenamefont {Kitaev}(2003)}]{Kitaev2003}%
  \BibitemOpen
  \bibfield  {author} {\bibinfo {author} {\bibfnamefont {A.}~\bibnamefont
  {Kitaev}},\ }\href {\doibase https://doi.org/10.1016/S0003-4916(02)00018-0}
  {\bibfield  {journal} {\bibinfo  {journal} {Annals of Physics}\ }\textbf
  {\bibinfo {volume} {303}},\ \bibinfo {pages} {2 } (\bibinfo {year}
  {2003})}\BibitemShut {NoStop}%
\bibitem [{\citenamefont {Banerjee}\ \emph {et~al.}(2016)\citenamefont
  {Banerjee}, \citenamefont {Bridges}, \citenamefont {Yan}, \citenamefont
  {Aczel}, \citenamefont {Li}, \citenamefont {Stone}, \citenamefont {Granroth},
  \citenamefont {Lumsden}, \citenamefont {Yiu}, \citenamefont {Knolle} \emph
  {et~al.}}]{Banerjee2016}%
  \BibitemOpen
  \bibfield  {author} {\bibinfo {author} {\bibfnamefont {A.}~\bibnamefont
  {Banerjee}}, \bibinfo {author} {\bibfnamefont {C.}~\bibnamefont {Bridges}},
  \bibinfo {author} {\bibfnamefont {J.-Q.}\ \bibnamefont {Yan}}, \bibinfo
  {author} {\bibfnamefont {A.}~\bibnamefont {Aczel}}, \bibinfo {author}
  {\bibfnamefont {L.}~\bibnamefont {Li}}, \bibinfo {author} {\bibfnamefont
  {M.}~\bibnamefont {Stone}}, \bibinfo {author} {\bibfnamefont
  {G.}~\bibnamefont {Granroth}}, \bibinfo {author} {\bibfnamefont
  {M.}~\bibnamefont {Lumsden}}, \bibinfo {author} {\bibfnamefont
  {Y.}~\bibnamefont {Yiu}}, \bibinfo {author} {\bibfnamefont {J.}~\bibnamefont
  {Knolle}},  \emph {et~al.},\ }\href@noop {} {\bibfield  {journal} {\bibinfo
  {journal} {Nature materials}\ }\textbf {\bibinfo {volume} {15}},\ \bibinfo
  {pages} {733} (\bibinfo {year} {2016})}\BibitemShut {NoStop}%
\bibitem [{\citenamefont {Banerjee}\ \emph {et~al.}(2017)\citenamefont
  {Banerjee}, \citenamefont {Yan}, \citenamefont {Knolle}, \citenamefont
  {Bridges}, \citenamefont {Stone}, \citenamefont {Lumsden}, \citenamefont
  {Mandrus}, \citenamefont {Tennant}, \citenamefont {Moessner},\ and\
  \citenamefont {Nagler}}]{Banerjee2017}%
  \BibitemOpen
  \bibfield  {author} {\bibinfo {author} {\bibfnamefont {A.}~\bibnamefont
  {Banerjee}}, \bibinfo {author} {\bibfnamefont {J.}~\bibnamefont {Yan}},
  \bibinfo {author} {\bibfnamefont {J.}~\bibnamefont {Knolle}}, \bibinfo
  {author} {\bibfnamefont {C.}~\bibnamefont {Bridges}}, \bibinfo {author}
  {\bibfnamefont {M.}~\bibnamefont {Stone}}, \bibinfo {author} {\bibfnamefont
  {M.}~\bibnamefont {Lumsden}}, \bibinfo {author} {\bibfnamefont
  {D.}~\bibnamefont {Mandrus}}, \bibinfo {author} {\bibfnamefont
  {D.}~\bibnamefont {Tennant}}, \bibinfo {author} {\bibfnamefont
  {R.}~\bibnamefont {Moessner}}, \ and\ \bibinfo {author} {\bibfnamefont
  {S.}~\bibnamefont {Nagler}},\ }\href {\doibase 10.1126/science.aah6015}
  {\bibfield  {journal} {\bibinfo  {journal} {Science}\ }\textbf {\bibinfo
  {volume} {356}},\ \bibinfo {pages} {1055} (\bibinfo {year}
  {2017})}\BibitemShut {NoStop}%
\bibitem [{\citenamefont {Ma}(2017)}]{Ma2017}%
  \BibitemOpen
  \bibfield  {author} {\bibinfo {author} {\bibfnamefont {N.}~\bibnamefont
  {Ma}},\ }\href {\doibase https://doi.org/10.1016/j.physb.2017.01.028}
  {\bibfield  {journal} {\bibinfo  {journal} {Physica B: Condensed Matter}\
  }\textbf {\bibinfo {volume} {512}},\ \bibinfo {pages} {100 } (\bibinfo {year}
  {2017})}\BibitemShut {NoStop}%
\bibitem [{\citenamefont {Zhang}\ \emph {et~al.}(2018)\citenamefont {Zhang},
  \citenamefont {Baker}, \citenamefont {Zhang}, \citenamefont {Wang},
  \citenamefont {Wang}, \citenamefont {Su}, \citenamefont {Zhu},\ and\
  \citenamefont {Pratt}}]{Zhang2018}%
  \BibitemOpen
  \bibfield  {author} {\bibinfo {author} {\bibfnamefont {B.}~\bibnamefont
  {Zhang}}, \bibinfo {author} {\bibfnamefont {P.}~\bibnamefont {Baker}},
  \bibinfo {author} {\bibfnamefont {Y.}~\bibnamefont {Zhang}}, \bibinfo
  {author} {\bibfnamefont {D.}~\bibnamefont {Wang}}, \bibinfo {author}
  {\bibfnamefont {Z.}~\bibnamefont {Wang}}, \bibinfo {author} {\bibfnamefont
  {S.}~\bibnamefont {Su}}, \bibinfo {author} {\bibfnamefont {D.}~\bibnamefont
  {Zhu}}, \ and\ \bibinfo {author} {\bibfnamefont {F.}~\bibnamefont {Pratt}},\
  }\href {\doibase 10.1021/jacs.7b11179} {\bibfield  {journal} {\bibinfo
  {journal} {Journal of the American Chemical Society}\ }\textbf {\bibinfo
  {volume} {140}},\ \bibinfo {pages} {122} (\bibinfo {year}
  {2018})}\BibitemShut {NoStop}%
\bibitem [{\citenamefont {Zheng}\ \emph {et~al.}(2017)\citenamefont {Zheng},
  \citenamefont {Ran}, \citenamefont {Li}, \citenamefont {Wang}, \citenamefont
  {Wang}, \citenamefont {Liu}, \citenamefont {Liu}, \citenamefont {Normand},
  \citenamefont {Wen},\ and\ \citenamefont {Yu}}]{Zheng2017}%
  \BibitemOpen
  \bibfield  {author} {\bibinfo {author} {\bibfnamefont {J.}~\bibnamefont
  {Zheng}}, \bibinfo {author} {\bibfnamefont {K.}~\bibnamefont {Ran}}, \bibinfo
  {author} {\bibfnamefont {T.}~\bibnamefont {Li}}, \bibinfo {author}
  {\bibfnamefont {J.}~\bibnamefont {Wang}}, \bibinfo {author} {\bibfnamefont
  {P.}~\bibnamefont {Wang}}, \bibinfo {author} {\bibfnamefont {B.}~\bibnamefont
  {Liu}}, \bibinfo {author} {\bibfnamefont {Z.-X.}\ \bibnamefont {Liu}},
  \bibinfo {author} {\bibfnamefont {B.}~\bibnamefont {Normand}}, \bibinfo
  {author} {\bibfnamefont {J.}~\bibnamefont {Wen}}, \ and\ \bibinfo {author}
  {\bibfnamefont {W.}~\bibnamefont {Yu}},\ }\href {\doibase
  10.1103/PhysRevLett.119.227208} {\bibfield  {journal} {\bibinfo  {journal}
  {Phys. Rev. Lett.}\ }\textbf {\bibinfo {volume} {119}},\ \bibinfo {pages}
  {227208} (\bibinfo {year} {2017})}\BibitemShut {NoStop}%
\bibitem [{\citenamefont {Do}\ \emph {et~al.}(2017)\citenamefont {Do},
  \citenamefont {Park}, \citenamefont {Yoshitake}, \citenamefont {Nasu},
  \citenamefont {Motome}, \citenamefont {Kwon}, \citenamefont {Adroja},
  \citenamefont {Voneshen}, \citenamefont {Kim}, \citenamefont {Jang} \emph
  {et~al.}}]{Do2017}%
  \BibitemOpen
  \bibfield  {author} {\bibinfo {author} {\bibfnamefont {S.-H.}\ \bibnamefont
  {Do}}, \bibinfo {author} {\bibfnamefont {S.-Y.}\ \bibnamefont {Park}},
  \bibinfo {author} {\bibfnamefont {J.}~\bibnamefont {Yoshitake}}, \bibinfo
  {author} {\bibfnamefont {J.}~\bibnamefont {Nasu}}, \bibinfo {author}
  {\bibfnamefont {Y.}~\bibnamefont {Motome}}, \bibinfo {author} {\bibfnamefont
  {Y.}~\bibnamefont {Kwon}}, \bibinfo {author} {\bibfnamefont {D.}~\bibnamefont
  {Adroja}}, \bibinfo {author} {\bibfnamefont {D.}~\bibnamefont {Voneshen}},
  \bibinfo {author} {\bibfnamefont {K.}~\bibnamefont {Kim}}, \bibinfo {author}
  {\bibfnamefont {T.-H.}\ \bibnamefont {Jang}},  \emph {et~al.},\ }\href@noop
  {} {\bibfield  {journal} {\bibinfo  {journal} {Nature Physics}\ }\textbf
  {\bibinfo {volume} {13}},\ \bibinfo {pages} {1079} (\bibinfo {year}
  {2017})}\BibitemShut {NoStop}%
\bibitem [{\citenamefont {Balz}\ \emph {et~al.}(2016)\citenamefont {Balz},
  \citenamefont {Lake}, \citenamefont {Reuther}, \citenamefont {Luetkens},
  \citenamefont {Sch{\"o}nemann}, \citenamefont {Herrmannsd{\"o}rfer},
  \citenamefont {Singh}, \citenamefont {Islam}, \citenamefont {Wheeler},
  \citenamefont {Rodriguez-Rivera} \emph {et~al.}}]{Balz2016}%
  \BibitemOpen
  \bibfield  {author} {\bibinfo {author} {\bibfnamefont {C.}~\bibnamefont
  {Balz}}, \bibinfo {author} {\bibfnamefont {B.}~\bibnamefont {Lake}}, \bibinfo
  {author} {\bibfnamefont {J.}~\bibnamefont {Reuther}}, \bibinfo {author}
  {\bibfnamefont {H.}~\bibnamefont {Luetkens}}, \bibinfo {author}
  {\bibfnamefont {R.}~\bibnamefont {Sch{\"o}nemann}}, \bibinfo {author}
  {\bibfnamefont {T.}~\bibnamefont {Herrmannsd{\"o}rfer}}, \bibinfo {author}
  {\bibfnamefont {Y.}~\bibnamefont {Singh}}, \bibinfo {author} {\bibfnamefont
  {A.}~\bibnamefont {Islam}}, \bibinfo {author} {\bibfnamefont
  {E.}~\bibnamefont {Wheeler}}, \bibinfo {author} {\bibfnamefont
  {J.}~\bibnamefont {Rodriguez-Rivera}},  \emph {et~al.},\ }\href@noop {}
  {\bibfield  {journal} {\bibinfo  {journal} {Nature Physics}\ }\textbf
  {\bibinfo {volume} {12}},\ \bibinfo {pages} {942} (\bibinfo {year}
  {2016})}\BibitemShut {NoStop}%
\bibitem [{\citenamefont {Harris}\ \emph {et~al.}(1997)\citenamefont {Harris},
  \citenamefont {Bramwell}, \citenamefont {McMorrow}, \citenamefont {Zeiske},\
  and\ \citenamefont {Godfrey}}]{Harris1997}%
  \BibitemOpen
  \bibfield  {author} {\bibinfo {author} {\bibfnamefont {M.~J.}\ \bibnamefont
  {Harris}}, \bibinfo {author} {\bibfnamefont {S.~T.}\ \bibnamefont
  {Bramwell}}, \bibinfo {author} {\bibfnamefont {D.~F.}\ \bibnamefont
  {McMorrow}}, \bibinfo {author} {\bibfnamefont {T.}~\bibnamefont {Zeiske}}, \
  and\ \bibinfo {author} {\bibfnamefont {K.~W.}\ \bibnamefont {Godfrey}},\
  }\href {\doibase 10.1103/PhysRevLett.79.2554} {\bibfield  {journal} {\bibinfo
   {journal} {Phys. Rev. Lett.}\ }\textbf {\bibinfo {volume} {79}},\ \bibinfo
  {pages} {2554} (\bibinfo {year} {1997})}\BibitemShut {NoStop}%
\bibitem [{\citenamefont {Fukazawa}\ \emph {et~al.}(2002)\citenamefont
  {Fukazawa}, \citenamefont {Melko}, \citenamefont {Higashinaka}, \citenamefont
  {Maeno},\ and\ \citenamefont {Gingras}}]{Fukazawa2002}%
  \BibitemOpen
  \bibfield  {author} {\bibinfo {author} {\bibfnamefont {H.}~\bibnamefont
  {Fukazawa}}, \bibinfo {author} {\bibfnamefont {R.~G.}\ \bibnamefont {Melko}},
  \bibinfo {author} {\bibfnamefont {R.}~\bibnamefont {Higashinaka}}, \bibinfo
  {author} {\bibfnamefont {Y.}~\bibnamefont {Maeno}}, \ and\ \bibinfo {author}
  {\bibfnamefont {M.~J.~P.}\ \bibnamefont {Gingras}},\ }\href {\doibase
  10.1103/PhysRevB.65.054410} {\bibfield  {journal} {\bibinfo  {journal} {Phys.
  Rev. B}\ }\textbf {\bibinfo {volume} {65}},\ \bibinfo {pages} {054410}
  (\bibinfo {year} {2002})}\BibitemShut {NoStop}%
\bibitem [{\citenamefont {Castelnovo}\ \emph {et~al.}(2008)\citenamefont
  {Castelnovo}, \citenamefont {Moessner},\ and\ \citenamefont
  {sondhi}}]{Castelnovo2008}%
  \BibitemOpen
  \bibfield  {author} {\bibinfo {author} {\bibfnamefont {C.}~\bibnamefont
  {Castelnovo}}, \bibinfo {author} {\bibfnamefont {R.}~\bibnamefont
  {Moessner}}, \ and\ \bibinfo {author} {\bibfnamefont {S.}~\bibnamefont
  {sondhi}},\ }\href@noop {} {\bibfield  {journal} {\bibinfo  {journal}
  {Nature}\ }\textbf {\bibinfo {volume} {451}},\ \bibinfo {pages} {42}
  (\bibinfo {year} {2008})}\BibitemShut {NoStop}%
\bibitem [{\citenamefont {Siddharthan}\ \emph {et~al.}(1999)\citenamefont
  {Siddharthan}, \citenamefont {Shastry}, \citenamefont {Ramirez},
  \citenamefont {Hayashi}, \citenamefont {Cava},\ and\ \citenamefont
  {Rosenkranz}}]{Siddharthan1999}%
  \BibitemOpen
  \bibfield  {author} {\bibinfo {author} {\bibfnamefont {R.}~\bibnamefont
  {Siddharthan}}, \bibinfo {author} {\bibfnamefont {B.~S.}\ \bibnamefont
  {Shastry}}, \bibinfo {author} {\bibfnamefont {A.~P.}\ \bibnamefont
  {Ramirez}}, \bibinfo {author} {\bibfnamefont {A.}~\bibnamefont {Hayashi}},
  \bibinfo {author} {\bibfnamefont {R.~J.}\ \bibnamefont {Cava}}, \ and\
  \bibinfo {author} {\bibfnamefont {S.}~\bibnamefont {Rosenkranz}},\ }\href
  {\doibase 10.1103/PhysRevLett.83.1854} {\bibfield  {journal} {\bibinfo
  {journal} {Phys. Rev. Lett.}\ }\textbf {\bibinfo {volume} {83}},\ \bibinfo
  {pages} {1854} (\bibinfo {year} {1999})}\BibitemShut {NoStop}%
\bibitem [{\citenamefont {Bramwell}\ and\ \citenamefont
  {Harris}(1998)}]{Bramwell1998}%
  \BibitemOpen
  \bibfield  {author} {\bibinfo {author} {\bibfnamefont {S.}~\bibnamefont
  {Bramwell}}\ and\ \bibinfo {author} {\bibfnamefont {M.}~\bibnamefont
  {Harris}},\ }\href {http://stacks.iop.org/0953-8984/10/i=14/a=002} {\bibfield
   {journal} {\bibinfo  {journal} {Journal of Physics: Condensed Matter}\
  }\textbf {\bibinfo {volume} {10}},\ \bibinfo {pages} {L215} (\bibinfo {year}
  {1998})}\BibitemShut {NoStop}%
\bibitem [{\citenamefont {Bramwell}\ and\ \citenamefont
  {Gingras}(2001)}]{Bramwell2001}%
  \BibitemOpen
  \bibfield  {author} {\bibinfo {author} {\bibfnamefont {S.}~\bibnamefont
  {Bramwell}}\ and\ \bibinfo {author} {\bibfnamefont {M.}~\bibnamefont
  {Gingras}},\ }\href {\doibase 10.1126/science.1064761} {\bibfield  {journal}
  {\bibinfo  {journal} {Science}\ }\textbf {\bibinfo {volume} {294}},\ \bibinfo
  {pages} {1495} (\bibinfo {year} {2001})},\ \Eprint
  {http://arxiv.org/abs/http://science.sciencemag.org/content/294/5546/1495.full.pdf}
  {http://science.sciencemag.org/content/294/5546/1495.full.pdf} \BibitemShut
  {NoStop}%
\bibitem [{\citenamefont {Greedan}(2006)}]{Greedan2006}%
  \BibitemOpen
  \bibfield  {author} {\bibinfo {author} {\bibfnamefont {J.~E.}\ \bibnamefont
  {Greedan}},\ }\href {\doibase https://doi.org/10.1016/j.jallcom.2004.12.084}
  {\bibfield  {journal} {\bibinfo  {journal} {Journal of Alloys and Compounds}\
  }\textbf {\bibinfo {volume} {408-412}},\ \bibinfo {pages} {444 } (\bibinfo
  {year} {2006})},\ \bibinfo {note} {proceedings of Rare Earths'04 in Nara,
  Japan}\BibitemShut {NoStop}%
\bibitem [{\citenamefont {Krey}\ \emph {et~al.}(2012)\citenamefont {Krey},
  \citenamefont {Legl}, \citenamefont {Dunsiger}, \citenamefont {Meven},
  \citenamefont {Gardner}, \citenamefont {Roper},\ and\ \citenamefont
  {Pfleiderer}}]{Krey2012}%
  \BibitemOpen
  \bibfield  {author} {\bibinfo {author} {\bibfnamefont {C.}~\bibnamefont
  {Krey}}, \bibinfo {author} {\bibfnamefont {S.}~\bibnamefont {Legl}}, \bibinfo
  {author} {\bibfnamefont {S.~R.}\ \bibnamefont {Dunsiger}}, \bibinfo {author}
  {\bibfnamefont {M.}~\bibnamefont {Meven}}, \bibinfo {author} {\bibfnamefont
  {J.~S.}\ \bibnamefont {Gardner}}, \bibinfo {author} {\bibfnamefont {J.~M.}\
  \bibnamefont {Roper}}, \ and\ \bibinfo {author} {\bibfnamefont
  {C.}~\bibnamefont {Pfleiderer}},\ }\href {\doibase
  10.1103/PhysRevLett.108.257204} {\bibfield  {journal} {\bibinfo  {journal}
  {Phys. Rev. Lett.}\ }\textbf {\bibinfo {volume} {108}},\ \bibinfo {pages}
  {257204} (\bibinfo {year} {2012})}\BibitemShut {NoStop}%
\bibitem [{\citenamefont {Hiroi}\ \emph {et~al.}(2003)\citenamefont {Hiroi},
  \citenamefont {Matsuhira}, \citenamefont {Takagi}, \citenamefont {Tayama},\
  and\ \citenamefont {Sakakibara}}]{Hiroi2003}%
  \BibitemOpen
  \bibfield  {author} {\bibinfo {author} {\bibfnamefont {Z.}~\bibnamefont
  {Hiroi}}, \bibinfo {author} {\bibfnamefont {K.}~\bibnamefont {Matsuhira}},
  \bibinfo {author} {\bibfnamefont {S.}~\bibnamefont {Takagi}}, \bibinfo
  {author} {\bibfnamefont {T.}~\bibnamefont {Tayama}}, \ and\ \bibinfo {author}
  {\bibfnamefont {T.}~\bibnamefont {Sakakibara}},\ }\href {\doibase
  10.1143/JPSJ.72.411} {\bibfield  {journal} {\bibinfo  {journal} {Journal of
  the Physical Society of Japan}\ }\textbf {\bibinfo {volume} {72}},\ \bibinfo
  {pages} {411} (\bibinfo {year} {2003})}\BibitemShut {NoStop}%
\bibitem [{\citenamefont {Cao}\ \emph {et~al.}(2009{\natexlab{a}})\citenamefont
  {Cao}, \citenamefont {Gukasov}, \citenamefont {Mirebeau}, \citenamefont
  {Bonville}, \citenamefont {Decorse},\ and\ \citenamefont
  {Dhalenne}}]{Cao2009}%
  \BibitemOpen
  \bibfield  {author} {\bibinfo {author} {\bibfnamefont {H.}~\bibnamefont
  {Cao}}, \bibinfo {author} {\bibfnamefont {A.}~\bibnamefont {Gukasov}},
  \bibinfo {author} {\bibfnamefont {I.}~\bibnamefont {Mirebeau}}, \bibinfo
  {author} {\bibfnamefont {P.}~\bibnamefont {Bonville}}, \bibinfo {author}
  {\bibfnamefont {C.}~\bibnamefont {Decorse}}, \ and\ \bibinfo {author}
  {\bibfnamefont {G.}~\bibnamefont {Dhalenne}},\ }\href {\doibase
  10.1103/PhysRevLett.103.056402} {\bibfield  {journal} {\bibinfo  {journal}
  {Phys. Rev. Lett.}\ }\textbf {\bibinfo {volume} {103}},\ \bibinfo {pages}
  {056402} (\bibinfo {year} {2009}{\natexlab{a}})}\BibitemShut {NoStop}%
\bibitem [{\citenamefont {Cao}\ \emph {et~al.}(2008)\citenamefont {Cao},
  \citenamefont {Gukasov}, \citenamefont {Mirebeau}, \citenamefont {Bonville},\
  and\ \citenamefont {Dhalenne.}}]{Cao2008a}%
  \BibitemOpen
  \bibfield  {author} {\bibinfo {author} {\bibfnamefont {H.}~\bibnamefont
  {Cao}}, \bibinfo {author} {\bibfnamefont {A.}~\bibnamefont {Gukasov}},
  \bibinfo {author} {\bibfnamefont {I.}~\bibnamefont {Mirebeau}}, \bibinfo
  {author} {\bibfnamefont {P.}~\bibnamefont {Bonville}}, \ and\ \bibinfo
  {author} {\bibfnamefont {G.}~\bibnamefont {Dhalenne.}},\ }\href {\doibase
  10.1103/PhysRevLett.101.196402} {\bibfield  {journal} {\bibinfo  {journal}
  {Phys. Rev. Lett.}\ }\textbf {\bibinfo {volume} {101}},\ \bibinfo {pages}
  {196402} (\bibinfo {year} {2008})}\BibitemShut {NoStop}%
\bibitem [{\citenamefont {Cao}\ \emph {et~al.}(2009{\natexlab{b}})\citenamefont
  {Cao}, \citenamefont {Gukasov}, \citenamefont {Mirebeau}, \citenamefont
  {Bonville}, \citenamefont {Decorse},\ and\ \citenamefont
  {Dhalenne}}]{Cao2008}%
  \BibitemOpen
  \bibfield  {author} {\bibinfo {author} {\bibfnamefont {H.}~\bibnamefont
  {Cao}}, \bibinfo {author} {\bibfnamefont {A.}~\bibnamefont {Gukasov}},
  \bibinfo {author} {\bibfnamefont {I.}~\bibnamefont {Mirebeau}}, \bibinfo
  {author} {\bibfnamefont {P.}~\bibnamefont {Bonville}}, \bibinfo {author}
  {\bibfnamefont {C.}~\bibnamefont {Decorse}}, \ and\ \bibinfo {author}
  {\bibfnamefont {G.}~\bibnamefont {Dhalenne}},\ }\href {\doibase
  10.1103/PhysRevLett.103.056402} {\bibfield  {journal} {\bibinfo  {journal}
  {Phys. Rev. Lett.}\ }\textbf {\bibinfo {volume} {103}},\ \bibinfo {pages}
  {056402} (\bibinfo {year} {2009}{\natexlab{b}})}\BibitemShut {NoStop}%
\bibitem [{\citenamefont {Isakov}\ \emph {et~al.}(2004)\citenamefont {Isakov},
  \citenamefont {Raman}, \citenamefont {Moessner},\ and\ \citenamefont
  {Sondhi}}]{Isakov2004}%
  \BibitemOpen
  \bibfield  {author} {\bibinfo {author} {\bibfnamefont {S.~V.}\ \bibnamefont
  {Isakov}}, \bibinfo {author} {\bibfnamefont {K.~S.}\ \bibnamefont {Raman}},
  \bibinfo {author} {\bibfnamefont {R.}~\bibnamefont {Moessner}}, \ and\
  \bibinfo {author} {\bibfnamefont {S.~L.}\ \bibnamefont {Sondhi}},\ }\href
  {\doibase 10.1103/PhysRevB.70.104418} {\bibfield  {journal} {\bibinfo
  {journal} {Phys. Rev. B}\ }\textbf {\bibinfo {volume} {70}},\ \bibinfo
  {pages} {104418} (\bibinfo {year} {2004})}\BibitemShut {NoStop}%
\bibitem [{\citenamefont {den Hertog}\ and\ \citenamefont
  {Gingras}(2000)}]{Hertog2000}%
  \BibitemOpen
  \bibfield  {author} {\bibinfo {author} {\bibfnamefont {B.~C.}\ \bibnamefont
  {den Hertog}}\ and\ \bibinfo {author} {\bibfnamefont {M.~J.~P.}\ \bibnamefont
  {Gingras}},\ }\href {\doibase 10.1103/PhysRevLett.84.3430} {\bibfield
  {journal} {\bibinfo  {journal} {Phys. Rev. Lett.}\ }\textbf {\bibinfo
  {volume} {84}},\ \bibinfo {pages} {3430} (\bibinfo {year}
  {2000})}\BibitemShut {NoStop}%
\bibitem [{\citenamefont {Becker}\ and\ \citenamefont
  {Felsche}(1987)}]{Becker1987}%
  \BibitemOpen
  \bibfield  {author} {\bibinfo {author} {\bibfnamefont {U.}~\bibnamefont
  {Becker}}\ and\ \bibinfo {author} {\bibfnamefont {J.}~\bibnamefont
  {Felsche}},\ }\href {\doibase https://doi.org/10.1016/0022-5088(87)90215-3}
  {\bibfield  {journal} {\bibinfo  {journal} {Journal of the Less Common
  Metals}\ }\textbf {\bibinfo {volume} {128}},\ \bibinfo {pages} {269 }
  (\bibinfo {year} {1987})}\BibitemShut {NoStop}%
\bibitem [{\citenamefont {Morosan}\ \emph {et~al.}(2008)\citenamefont
  {Morosan}, \citenamefont {Fleitman}, \citenamefont {Huang}, \citenamefont
  {Lynn}, \citenamefont {Chen}, \citenamefont {Ke}, \citenamefont {Dahlberg},
  \citenamefont {Schiffer}, \citenamefont {Craley},\ and\ \citenamefont
  {Cava}}]{Morosan2008}%
  \BibitemOpen
  \bibfield  {author} {\bibinfo {author} {\bibfnamefont {E.}~\bibnamefont
  {Morosan}}, \bibinfo {author} {\bibfnamefont {J.~A.}\ \bibnamefont
  {Fleitman}}, \bibinfo {author} {\bibfnamefont {Q.}~\bibnamefont {Huang}},
  \bibinfo {author} {\bibfnamefont {J.~W.}\ \bibnamefont {Lynn}}, \bibinfo
  {author} {\bibfnamefont {Y.}~\bibnamefont {Chen}}, \bibinfo {author}
  {\bibfnamefont {X.}~\bibnamefont {Ke}}, \bibinfo {author} {\bibfnamefont
  {M.~L.}\ \bibnamefont {Dahlberg}}, \bibinfo {author} {\bibfnamefont
  {P.}~\bibnamefont {Schiffer}}, \bibinfo {author} {\bibfnamefont {C.~R.}\
  \bibnamefont {Craley}}, \ and\ \bibinfo {author} {\bibfnamefont {R.~J.}\
  \bibnamefont {Cava}},\ }\href {\doibase 10.1103/PhysRevB.77.224423}
  {\bibfield  {journal} {\bibinfo  {journal} {Phys. Rev. B}\ }\textbf {\bibinfo
  {volume} {77}},\ \bibinfo {pages} {224423} (\bibinfo {year}
  {2008})}\BibitemShut {NoStop}%
\bibitem [{\citenamefont {Ke}\ \emph {et~al.}(2008)\citenamefont {Ke},
  \citenamefont {Dahlberg}, \citenamefont {Morosan}, \citenamefont {Fleitman},
  \citenamefont {Cava},\ and\ \citenamefont {Schiffer}}]{Ke2008}%
  \BibitemOpen
  \bibfield  {author} {\bibinfo {author} {\bibfnamefont {X.}~\bibnamefont
  {Ke}}, \bibinfo {author} {\bibfnamefont {M.~L.}\ \bibnamefont {Dahlberg}},
  \bibinfo {author} {\bibfnamefont {E.}~\bibnamefont {Morosan}}, \bibinfo
  {author} {\bibfnamefont {J.~A.}\ \bibnamefont {Fleitman}}, \bibinfo {author}
  {\bibfnamefont {R.~J.}\ \bibnamefont {Cava}}, \ and\ \bibinfo {author}
  {\bibfnamefont {P.}~\bibnamefont {Schiffer}},\ }\href {\doibase
  10.1103/PhysRevB.78.104411} {\bibfield  {journal} {\bibinfo  {journal} {Phys.
  Rev. B}\ }\textbf {\bibinfo {volume} {78}},\ \bibinfo {pages} {104411}
  (\bibinfo {year} {2008})}\BibitemShut {NoStop}%
\bibitem [{\citenamefont {Ghosh}\ \emph {et~al.}(1998)\citenamefont {Ghosh},
  \citenamefont {Jana}, \citenamefont {Ghosh},\ and\ \citenamefont
  {Wanklyn}}]{Ghosh1998}%
  \BibitemOpen
  \bibfield  {author} {\bibinfo {author} {\bibfnamefont {M.}~\bibnamefont
  {Ghosh}}, \bibinfo {author} {\bibfnamefont {S.}~\bibnamefont {Jana}},
  \bibinfo {author} {\bibfnamefont {D.}~\bibnamefont {Ghosh}}, \ and\ \bibinfo
  {author} {\bibfnamefont {B.}~\bibnamefont {Wanklyn}},\ }\href {\doibase
  https://doi.org/10.1016/S0038-1098(98)00161-6} {\bibfield  {journal}
  {\bibinfo  {journal} {Solid State Communications}\ }\textbf {\bibinfo
  {volume} {107}},\ \bibinfo {pages} {113 } (\bibinfo {year}
  {1998})}\BibitemShut {NoStop}%
\bibitem [{\citenamefont {Snyder}\ \emph {et~al.}(2001)\citenamefont {Snyder},
  \citenamefont {Slusky}, \citenamefont {Cava},\ and\ \citenamefont
  {Schiffer}}]{Snyder2001}%
  \BibitemOpen
  \bibfield  {author} {\bibinfo {author} {\bibfnamefont {J.}~\bibnamefont
  {Snyder}}, \bibinfo {author} {\bibfnamefont {J.}~\bibnamefont {Slusky}},
  \bibinfo {author} {\bibfnamefont {R.}~\bibnamefont {Cava}}, \ and\ \bibinfo
  {author} {\bibfnamefont {P.}~\bibnamefont {Schiffer}},\ }\href@noop {}
  {\bibfield  {journal} {\bibinfo  {journal} {Nature}\ }\textbf {\bibinfo
  {volume} {413}},\ \bibinfo {pages} {48} (\bibinfo {year} {2001})}\BibitemShut
  {NoStop}%
\bibitem [{\citenamefont {Snyder}\ \emph {et~al.}(2004)\citenamefont {Snyder},
  \citenamefont {Ueland}, \citenamefont {Slusky}, \citenamefont {Karunadasa},
  \citenamefont {Cava},\ and\ \citenamefont {Schiffer}}]{Snyder2004}%
  \BibitemOpen
  \bibfield  {author} {\bibinfo {author} {\bibfnamefont {J.}~\bibnamefont
  {Snyder}}, \bibinfo {author} {\bibfnamefont {B.~G.}\ \bibnamefont {Ueland}},
  \bibinfo {author} {\bibfnamefont {J.~S.}\ \bibnamefont {Slusky}}, \bibinfo
  {author} {\bibfnamefont {H.}~\bibnamefont {Karunadasa}}, \bibinfo {author}
  {\bibfnamefont {R.~J.}\ \bibnamefont {Cava}}, \ and\ \bibinfo {author}
  {\bibfnamefont {P.}~\bibnamefont {Schiffer}},\ }\href {\doibase
  10.1103/PhysRevB.69.064414} {\bibfield  {journal} {\bibinfo  {journal} {Phys.
  Rev. B}\ }\textbf {\bibinfo {volume} {69}},\ \bibinfo {pages} {064414}
  (\bibinfo {year} {2004})}\BibitemShut {NoStop}%
\bibitem [{\citenamefont {Ehlers}\ \emph {et~al.}(2004)\citenamefont {Ehlers},
  \citenamefont {Cornelius}, \citenamefont {Fennell}, \citenamefont {Koza},
  \citenamefont {Bramwell},\ and\ \citenamefont {Gardner}}]{Ehlers2004}%
  \BibitemOpen
  \bibfield  {author} {\bibinfo {author} {\bibfnamefont {G.}~\bibnamefont
  {Ehlers}}, \bibinfo {author} {\bibfnamefont {A.}~\bibnamefont {Cornelius}},
  \bibinfo {author} {\bibfnamefont {T.}~\bibnamefont {Fennell}}, \bibinfo
  {author} {\bibfnamefont {M.}~\bibnamefont {Koza}}, \bibinfo {author}
  {\bibfnamefont {S.}~\bibnamefont {Bramwell}}, \ and\ \bibinfo {author}
  {\bibfnamefont {J.}~\bibnamefont {Gardner}},\ }\href@noop {} {\bibfield
  {journal} {\bibinfo  {journal} {Journal of Physics: Condensed Matter}\
  }\textbf {\bibinfo {volume} {16}},\ \bibinfo {pages} {S635} (\bibinfo {year}
  {2004})}\BibitemShut {NoStop}%
\bibitem [{\citenamefont {Sheldrick}(2015)}]{Sheldrick2015}%
  \BibitemOpen
  \bibfield  {author} {\bibinfo {author} {\bibfnamefont {G.~M.}\ \bibnamefont
  {Sheldrick}},\ }\href {\doibase 10.1107/S2053229614024218} {\bibfield
  {journal} {\bibinfo  {journal} {Acta Crystallographica Section C}\ }\textbf
  {\bibinfo {volume} {71}},\ \bibinfo {pages} {3} (\bibinfo {year}
  {2015})}\BibitemShut {NoStop}%
\bibitem [{\citenamefont
  {Rodr{\'{\i}}guez-Carvajal}(1993)}]{Rodriguez-Carvajal1993}%
  \BibitemOpen
  \bibfield  {author} {\bibinfo {author} {\bibfnamefont {J.}~\bibnamefont
  {Rodr{\'{\i}}guez-Carvajal}},\ }\href {\doibase 10.1016/0921-4526(93)90108-I}
  {\bibfield  {journal} {\bibinfo  {journal} {Physica B: Condensed Matter}\
  }\textbf {\bibinfo {volume} {192}},\ \bibinfo {pages} {55} (\bibinfo {year}
  {1993})}\BibitemShut {NoStop}%
\bibitem [{\citenamefont {Finger}\ \emph {et~al.}(1994)\citenamefont {Finger},
  \citenamefont {Cox},\ and\ \citenamefont {Jephcoat}}]{Finger1994}%
  \BibitemOpen
  \bibfield  {author} {\bibinfo {author} {\bibfnamefont {L.~W.}\ \bibnamefont
  {Finger}}, \bibinfo {author} {\bibfnamefont {D.~E.}\ \bibnamefont {Cox}}, \
  and\ \bibinfo {author} {\bibfnamefont {A.~P.}\ \bibnamefont {Jephcoat}},\
  }\href {\doibase 10.1107/S0021889894004218} {\bibfield  {journal} {\bibinfo
  {journal} {J. Appl. Cryst.}\ }\textbf {\bibinfo {volume} {27}},\ \bibinfo
  {pages} {892} (\bibinfo {year} {1994})}\BibitemShut {NoStop}%
\bibitem [{\citenamefont {Wills}(2000)}]{Wills2000}%
  \BibitemOpen
  \bibfield  {author} {\bibinfo {author} {\bibfnamefont {A.}~\bibnamefont
  {Wills}},\ }\href {\doibase https://doi.org/10.1016/S0921-4526(99)01722-6}
  {\bibfield  {journal} {\bibinfo  {journal} {Physica B: Condensed Matter}\
  }\textbf {\bibinfo {volume} {276-278}},\ \bibinfo {pages} {680 } (\bibinfo
  {year} {2000})}\BibitemShut {NoStop}%
\bibitem [{\citenamefont {Campbell}\ \emph {et~al.}(2006)\citenamefont
  {Campbell}, \citenamefont {Stokes}, \citenamefont {Tanner},\ and\
  \citenamefont {Hatch}}]{Campbell2006}%
  \BibitemOpen
  \bibfield  {author} {\bibinfo {author} {\bibfnamefont {B.~J.}\ \bibnamefont
  {Campbell}}, \bibinfo {author} {\bibfnamefont {H.~T.}\ \bibnamefont
  {Stokes}}, \bibinfo {author} {\bibfnamefont {D.~E.}\ \bibnamefont {Tanner}},
  \ and\ \bibinfo {author} {\bibfnamefont {D.~M.}\ \bibnamefont {Hatch}},\
  }\href {\doibase 10.1107/S0021889806014075} {\bibfield  {journal} {\bibinfo
  {journal} {Journal of Applied Crystallography}\ }\textbf {\bibinfo {volume}
  {39}},\ \bibinfo {pages} {607} (\bibinfo {year} {2006})}\BibitemShut
  {NoStop}%
\bibitem [{Mom(2011)}]{Momma2011}%
  \BibitemOpen
  \href@noop {} {\bibfield  {journal} {\bibinfo  {journal} {Journal of Applied
  Crystallography}\ }\textbf {\bibinfo {volume} {44}},\ \bibinfo {pages} {1272}
  (\bibinfo {year} {2011})}\BibitemShut {NoStop}%
\bibitem [{\citenamefont {Shannon}(1976)}]{Shannon1976}%
  \BibitemOpen
  \bibfield  {author} {\bibinfo {author} {\bibfnamefont {R.~D.}\ \bibnamefont
  {Shannon}},\ }\href {\doibase 10.1107/S0567739476001551} {\bibfield
  {journal} {\bibinfo  {journal} {Acta Crystallographica Section A}\ }\textbf
  {\bibinfo {volume} {32}},\ \bibinfo {pages} {751} (\bibinfo {year}
  {1976})}\BibitemShut {NoStop}%
\bibitem [{\citenamefont {Moon}(1982)}]{Moon1982}%
  \BibitemOpen
  \bibfield  {author} {\bibinfo {author} {\bibfnamefont {R.}~\bibnamefont
  {Moon}},\ }\href {\doibase 10.1051/jphyscol:1982727} {\bibfield  {journal}
  {\bibinfo  {journal} {{Journal de Physique Colloques}}\ }\textbf {\bibinfo
  {volume} {43}},\ \bibinfo {pages} {C7} (\bibinfo {year} {1982})}\BibitemShut
  {NoStop}%
\bibitem [{SM()}]{SM}%
  \BibitemOpen
  \href@noop {} {}\bibinfo {note} {See Supplemental Material at}\BibitemShut
  {NoStop}%
\bibitem [{\citenamefont {Balents}(2010)}]{Balents2010}%
  \BibitemOpen
  \bibfield  {author} {\bibinfo {author} {\bibfnamefont {L.}~\bibnamefont
  {Balents}},\ }\href@noop {} {\bibfield  {journal} {\bibinfo  {journal}
  {Nature}\ }\textbf {\bibinfo {volume} {464}},\ \bibinfo {pages} {199}
  (\bibinfo {year} {2010})}\BibitemShut {NoStop}%
\bibitem [{\citenamefont {Stryjewski}\ and\ \citenamefont
  {Giordano}(1977)}]{Stryjewski1977}%
  \BibitemOpen
  \bibfield  {author} {\bibinfo {author} {\bibfnamefont {E.}~\bibnamefont
  {Stryjewski}}\ and\ \bibinfo {author} {\bibfnamefont {N.}~\bibnamefont
  {Giordano}},\ }\href {\doibase 10.1080/00018737700101433} {\bibfield
  {journal} {\bibinfo  {journal} {Advances in Physics}\ }\textbf {\bibinfo
  {volume} {26}},\ \bibinfo {pages} {487} (\bibinfo {year} {1977})}\BibitemShut
  {NoStop}%
\bibitem [{\citenamefont {Kitano}\ and\ \citenamefont
  {Nagamiya}(1964)}]{Kitano1964}%
  \BibitemOpen
  \bibfield  {author} {\bibinfo {author} {\bibfnamefont {Y.}~\bibnamefont
  {Kitano}}\ and\ \bibinfo {author} {\bibfnamefont {T.}~\bibnamefont
  {Nagamiya}},\ }\href@noop {} {\bibfield  {journal} {\bibinfo  {journal}
  {Progress of Theoretical Physics}\ }\textbf {\bibinfo {volume} {31}},\
  \bibinfo {pages} {1} (\bibinfo {year} {1964})}\BibitemShut {NoStop}%
\bibitem [{\citenamefont {Deppe}\ \emph {et~al.}(2012)\citenamefont {Deppe},
  \citenamefont {Lausberg}, \citenamefont {Weickert}, \citenamefont {Brando},
  \citenamefont {Skourski}, \citenamefont {Caroca-Canales}, \citenamefont
  {Geibel},\ and\ \citenamefont {Steglich}}]{Deppe2012}%
  \BibitemOpen
  \bibfield  {author} {\bibinfo {author} {\bibfnamefont {M.}~\bibnamefont
  {Deppe}}, \bibinfo {author} {\bibfnamefont {S.}~\bibnamefont {Lausberg}},
  \bibinfo {author} {\bibfnamefont {F.}~\bibnamefont {Weickert}}, \bibinfo
  {author} {\bibfnamefont {M.}~\bibnamefont {Brando}}, \bibinfo {author}
  {\bibfnamefont {Y.}~\bibnamefont {Skourski}}, \bibinfo {author}
  {\bibfnamefont {N.}~\bibnamefont {Caroca-Canales}}, \bibinfo {author}
  {\bibfnamefont {C.}~\bibnamefont {Geibel}}, \ and\ \bibinfo {author}
  {\bibfnamefont {F.}~\bibnamefont {Steglich}},\ }\href {\doibase
  10.1103/PhysRevB.85.060401} {\bibfield  {journal} {\bibinfo  {journal} {Phys.
  Rev. B}\ }\textbf {\bibinfo {volume} {85}},\ \bibinfo {pages} {060401}
  (\bibinfo {year} {2012})}\BibitemShut {NoStop}%
\bibitem [{\citenamefont {Medrano}\ \emph {et~al.}(2017)\citenamefont
  {Medrano}, \citenamefont {Freitas}, \citenamefont {Passamani}, \citenamefont
  {Pinheiro}, \citenamefont {Baggio-Saitovitch}, \citenamefont {Continentino},\
  and\ \citenamefont {Sanchez}}]{Medrano2017}%
  \BibitemOpen
  \bibfield  {author} {\bibinfo {author} {\bibfnamefont {C.~P.~C.}\
  \bibnamefont {Medrano}}, \bibinfo {author} {\bibfnamefont {D.~C.}\
  \bibnamefont {Freitas}}, \bibinfo {author} {\bibfnamefont {E.~C.}\
  \bibnamefont {Passamani}}, \bibinfo {author} {\bibfnamefont {C.~B.}\
  \bibnamefont {Pinheiro}}, \bibinfo {author} {\bibfnamefont {E.}~\bibnamefont
  {Baggio-Saitovitch}}, \bibinfo {author} {\bibfnamefont {M.~A.}\ \bibnamefont
  {Continentino}}, \ and\ \bibinfo {author} {\bibfnamefont {D.~R.}\
  \bibnamefont {Sanchez}},\ }\href {\doibase 10.1103/PhysRevB.95.214419}
  {\bibfield  {journal} {\bibinfo  {journal} {Phys. Rev. B}\ }\textbf {\bibinfo
  {volume} {95}},\ \bibinfo {pages} {214419} (\bibinfo {year}
  {2017})}\BibitemShut {NoStop}%
\bibitem [{\citenamefont {Sibille}\ \emph {et~al.}(2016)\citenamefont
  {Sibille}, \citenamefont {Lhotel}, \citenamefont {Hatnean}, \citenamefont
  {Balakrishnan}, \citenamefont {F\aa{}k}, \citenamefont {Gauthier},
  \citenamefont {Fennell},\ and\ \citenamefont {Kenzelmann}}]{Sibille2016}%
  \BibitemOpen
  \bibfield  {author} {\bibinfo {author} {\bibfnamefont {R.}~\bibnamefont
  {Sibille}}, \bibinfo {author} {\bibfnamefont {E.}~\bibnamefont {Lhotel}},
  \bibinfo {author} {\bibfnamefont {M.~C.}\ \bibnamefont {Hatnean}}, \bibinfo
  {author} {\bibfnamefont {G.}~\bibnamefont {Balakrishnan}}, \bibinfo {author}
  {\bibfnamefont {B.}~\bibnamefont {F\aa{}k}}, \bibinfo {author} {\bibfnamefont
  {N.}~\bibnamefont {Gauthier}}, \bibinfo {author} {\bibfnamefont
  {T.}~\bibnamefont {Fennell}}, \ and\ \bibinfo {author} {\bibfnamefont
  {M.}~\bibnamefont {Kenzelmann}},\ }\href {\doibase
  10.1103/PhysRevB.94.024436} {\bibfield  {journal} {\bibinfo  {journal} {Phys.
  Rev. B}\ }\textbf {\bibinfo {volume} {94}},\ \bibinfo {pages} {024436}
  (\bibinfo {year} {2016})}\BibitemShut {NoStop}%
\end{thebibliography}
\end{document}


\preprint{APS/123-QED}

\title{Supplemental Material on: \\ Local-Ising type magnetic order and metamagnetism in the rare-earth pyrogermanate Er$_2$Ge$_2$O$_7$}

\author{K.M. Taddei}
\email[corresponding author ]{taddeikm@ornl.gov}
\affiliation{Neutron Scattering Division, Oak Ridge National Laboratory, Oak Ridge, TN 37831}
\author{L. Sanjeewa}
\affiliation{Materials Science and Technology Division, Oak Ridge National Laboratory, Oak Ridge, TN 37831}
\affiliation{Department of Chemistry, Clemson University, Clemson, SC 29634}
\author{J.W. Kolis}
\affiliation{Department of Chemistry, Clemson University, Clemson, SC 29634}
\author{A.S. Sefat}
\affiliation{Materials Science and Technology Division, Oak Ridge National Laboratory, Oak Ridge, TN 37831}
\author{C. de la Cruz}
\affiliation{Neutron Scattering Division, Oak Ridge National Laboratory, Oak Ridge, TN 37831}
\author{D.M. Pajerowski}
\affiliation{Neutron Scattering Division, Oak Ridge National Laboratory, Oak Ridge, TN 37831}

\date{\today}

\maketitle

\section{\label{sec:sym} Symmetry analysis of magnetic order in Er$_2$Ge$_2$O$_7$}

Here we show the results of the symmetry analysis of space group symmetry $P4_12_12$ for ordering vector $k = (0,0,0)$. The five irreducible representations ($\Gamma$) are shown explicitly with their constituent basis vectors ($\bfpsi$). Each $\bfpsi$ is shown decomposed into its magnetic components along the crystallographic axes \textit{a}, \textit{b} and \textit{c} for each Er position in the unit cell. The Er positions 1-8 are at positions $( .876,~ .353,~ .136)$, $( .376,~ .147,~ .615)$, $( .624,~ .853,~ .115)$, $( .124,~ .647,~ .636)$, $( .647,~ .124,~ .365)$, $( .147,~ .376,~ .386)$, $( .853,~ .624,~ .886)$ and $( .353,~ .876,~ .866)$ respectively. The $\Gamma_5$ space group symmetries listed in the main text are composed of selections and linear combinations of the $\bfpsi$ listed here. $P2_{1}^{'}2_1^{1}2$ consists of $\bfpsi_{13}$, $\bfpsi_{14}$, $\bfpsi_{15}$, $\bfpsi_{16}$, $\bfpsi_{17}$ and $\bfpsi_{18}$. $C22^{'}2_{1}^{'}$ consists of ($\bfpsi_{13}-\bfpsi_{19}$), ($\bfpsi_{16}+\bfpsi_{22}$), ($\bfpsi_{14}+\bfpsi_{20}$), ($-\bfpsi_{17}+\bfpsi_{23}$), ($\bfpsi_{15}-\bfpsi_{21}$) and ($-\bfpsi_{18}+\bfpsi_{24}$). $\Gamma$ and $\bfpsi$ were generated using the SARAh rep analysis software \cite{Wills2000}. Assignment of $\bfpsi$ to magnetic space group symmetries was determined using the ISODISTORT suite \cite{Campbell2006}.

\clearpage

\begin{table}[t]
\caption{Basis vectors ($\bfpsi$) for $\Gamma_1$ and $\Gamma_2$}
\begin{tabular*}{0.7\textwidth}{c@{\extracolsep{\fill}}cc|rrr|rrrr|rrr}
  $\Gamma$  &  $\bfpsi$  &  Er & \multicolumn{3}{c}{ components} & &$\Gamma$  &  $\bfpsi$  &  Er & \multicolumn{3}{c}{ components}\\
      &      &             &$m_{a}$ & $m_{b}$ & $m_{c}$ &       &      &     &        &$m_{a}$ & $m_{b}$ & $m_{c}$  \\
\hline
$\Gamma_{1}$ & $\bfpsi_{1}$ &     1 &     1 &     0 &      0  & &$\Gamma_{2}$ & $\bfpsi_{4}$ & 1 & 1  & 0  &  0  \\
             &              &     2 &     1 &     0 &     0  &  &            &              & 2 & 1 & 0 & 0   \\
             &              &      3 &     -1 &      0 &      0 &    &      &              & 3 & -1 & 0 & 0     \\
             &              &      4 &     -1 &      0 &      0 &    &      &              & 4 & -1 & 0 & 0     \\
             &              &      5 &      0 &     -1 &      0 &    &        &              & 5 & 0 & 1 & 0   \\
             &              &      6 &      0 &      1 &      0 &  & &              & 6 & 0 & -1 & 0        \\
             &              &      7 &      0 &     -1 &      0 & &   &              & 7 & 0 & 1 & 0      \\
             &              &      8 &      0 &      1 &      0 & &  &              & 8 & 0 & -1 & 0     \\
             & $\bfpsi_{2}$ &      1 &      0 &     1 &    0 & &  & $\bfpsi_{5}$ & 1 & 0  & 1  &  0     \\
             &              &      2 &      0 &     -1 &      0 & & &              & 2 & 0 & -1 & 0     \\
             &              &      3 &      0 &      1 &      0 & &   &              & 3 & 0 & 1 & 0     \\
             &              &      4 &      0 &     -1 &      0 & &  &              & 4 & 0 & -1 & 0    \\
             &              &      5 &     -1 &      0 &      0 & &  &              & 5 & 1 & 0 & 0         \\
             &              &      6 &     -1 &      0 &      0 & &   &              & 6 & 1 & 0 & 0         \\
             &              &      7 &      1 &      0 &      0 & &    &              & 7 & -1 & 0 & 0      \\
             &              &      8 &      1 &      0 &      0 & &    &              & 8 & -1 & 0 & 0       \\
             & $\bfpsi_{3}$ &      1 &   0 &   0 &  1 &   & & $\bfpsi_{6}$ & 1 & 0  & 0  & 1         \\
             &              &      2 &      0 &      0 &     -1 &  &  &  & 2 & 0  & 0  & -1      \\
             &              &      3 &      0 &      0 &     -1 &   &  &  & 3 & 0  & 0  &- 1      \\
             &              &      4 &      0 &      0 &      1 &  &  &  & 4 & 0  & 0  & 1       \\
             &              &      5 &      0 &      0 &     -1 &   &  &  & 5 & 0  & 0  & 1    \\
             &              &      6 &      0 &      0 &      1 &   &  &  & 6 & 0  & 0  & -1     \\
             &              &      7 &      0 &      0 &      1 &    &  &  & 7 & 0  & 0  & -1    \\
             &              &      8 &      0 &      0 &     -1 &    &  &  & 8 & 0  & 0  & 1    \\

\end{tabular*}
\end{table}

\begin{table}[b]
\caption{Basis vectors ($\bfpsi$) for irreps $\Gamma_3$ and $\Gamma_4$}
\begin{tabular*}{0.7\textwidth}{c@{\extracolsep{\fill}}cc|rrr|rrrr|rrr}
  $\Gamma$  &  $\bfpsi$  &  Er & \multicolumn{3}{c}{ components} & &$\Gamma$  &  $\bfpsi$  &  Er & \multicolumn{3}{c}{ components}\\
      &      &             &$m_{a}$ & $m_{b}$ & $m_{c}$ &       &      &     &        &$m_{a}$ & $m_{b}$ & $m_{c}$  \\
\hline
$\Gamma_{3}$ & $\bfpsi_{7}$ &     1 &     1 &     0 &      0  & &$\Gamma_{4}$ & $\bfpsi_{10}$ & 1 & 1  & 0  &  0  \\
             &              &     2 &    -1 &     0 &     0  &  &            &    & 2 & -1 & 0 & 0   \\
             &              &      3 &    1 &      0 &      0 &    &      &       & 3 & 1 & 0 & 0     \\
             &              &      4 &    -1 &      0 &      0 &    &      &      & 4 & -1 & 0 & 0     \\
             &              &      5 &      0 &     1 &      0 &    &        &    & 5 & 0 & -1 & 0   \\
             &              &      6 &      0 &     1 &      0 &  & &             & 6 & 0 & -1 & 0        \\
             &              &      7 &      0 &    -1 &      0 & &   &            & 7 & 0 & 1 & 0      \\
             &              &      8 &      0 &    -1 &      0 & &  &             & 8 & 0 & 1 & 0     \\
             & $\bfpsi_{8}$ &      1 &      0 &    1 &    0 & &  & $\bfpsi_{11}$ & 1 & 0  & 1  &  0     \\
             &              &      2 &      0 &    1 &      0 & & &              & 2 & 0 & 1 & 0     \\
             &              &      3 &      0 &   -1 &      0 & &   &            & 3 & 0 & -1 & 0     \\
             &              &      4 &      0 &   -1 &      0 & &  &             & 4 & 0 & -1 & 0    \\
             &              &      5 &     1 &    0 &      0 & &  &              & 5 & -1 & 0 & 0         \\
             &              &      6 &    -1 &    0 &      0 & &   &             & 6 & 1 & 0 & 0         \\
             &              &      7 &     1 &      0 &      0 & &    &          & 7 & -1 & 0 & 0      \\
             &              &      8 &    -1 &      0 &      0 & &    &          & 8 & 1 & 0 & 0       \\
             & $\bfpsi_{9}$ &      1 &   0 &   0 &  1 &   & & $\bfpsi_{12}$ & 1 & 0  & 0  & 1         \\
             &              &      2 &      0 &      0 &     1 &  &  &  & 2 & 0  & 0  & 1      \\
             &              &      3 &      0 &      0 &     1 &   &  &  & 3 & 0  & 0  & 1      \\
             &              &      4 &      0 &      0 &     1 &  &  &  & 4 & 0  & 0  & 1       \\
             &              &      5 &      0 &      0 &     1 &   &  &  & 5 & 0  & 0 & -1    \\
             &              &      6 &      0 &      0 &     1 &   &  &  & 6 & 0  & 0 & -1     \\
             &              &      7 &      0 &      0 &     1 &    &  &  & 7 & 0  & 0& -1    \\
             &              &      8 &      0 &      0 &     1 &    &  &  & 8 & 0  & 0 & -1    \\

\end{tabular*}
\end{table}

\begin{table}[h]
\caption{First six basis vectors ($\bfpsi$) for irrep $\Gamma_5$}
\begin{tabular*}{0.7\textwidth}{c@{\extracolsep{\fill}}cc|rrr|rrrr|rrr}
  $\Gamma$  &  $\bfpsi$  &  Er & \multicolumn{3}{c}{components} & & $\Gamma$  &  $\bfpsi$  &  Er & \multicolumn{3}{c}{ components}\\
      &      &             &$m_{a}$ & $m_{b}$ & $m_{c}$ &       &      &     &        &$m_{a}$ & $m_{b}$ & $m_{c}$  \\
\hline
$\Gamma_{5}$ & $\bfpsi_{13}$ &     1 &     1 &     0 &   0  & & & $\bfpsi_{16}$ & 1 & 0  & 0  &  0  \\
             &              &     2 &    1 &     0 &     0  &  &            &    & 2 & 0 & 0 & 0   \\
             &              &      3 &    1 &      0 &      0 &    &      &       & 3 & 0 & 0 & 0     \\
             &              &      4 &    1 &      0 &      0 &    &      &      & 4 & 0 & 0 & 0     \\
             &              &      5 &      0 &     0 &      0 &    &        &    & 5 & 0 & -1 & 0   \\
             &              &      6 &      0 &    0 &      0 &  & &             & 6 & 0 & 1 & 0        \\
             &              &      7 &      0 &    0 &      0 & &   &            & 7 & 0 & -1 & 0      \\
             &              &      8 &      0 &    0 &      0 & &  &             & 8 & 0 & 1 & 0     \\
             & $\bfpsi_{14}$ &      1 &      0 &    1 &    0 & &  & $\bfpsi_{17}$ & 1 & 0  & 0  &  0     \\
             &              &      2 &      0 &    -1 &      0 & & &              & 2 & 0 & 0 & 0     \\
             &              &      3 &      0 &   -1 &      0 & &   &            & 3 & 0 & 0 & 0     \\
             &              &      4 &      0 &   1 &      0 & &  &             & 4 & 0 & 0 & 0    \\
             &              &      5 &     0 &    0 &      0 & &  &              & 5 & -1 & 0 & 0         \\
             &              &      6 &    0 &    0 &      0 & &   &             & 6 & -1 & 0 & 0         \\
             &              &      7 &     0 &      0 &      0 & &    &          & 7 & -1 & 0 & 0      \\
             &              &      8 &    0 &      0 &      0 & &    &          & 8 & -1 & 0 & 0       \\
             & $\bfpsi_{15}$ &      1 &   0 &   0 &  1 &   & & $\bfpsi_{18}$ & 1 & 0  & 0  & 0         \\
             &              &      2 &      0 &   0 &-1 &  &  &  & 2 & 0  & 0  &0      \\
             &              &      3 &      0 &   0 &  1 &   &  &  & 3 & 0  & 0  & 0      \\
             &              &      4 &      0 &   0 & -1 &  &  &  & 4 & 0  & 0  & 0       \\
             &              &      5 &      0 &   0 & 0 &   &  &  & 5 & 0  & 0 & -1    \\
             &              &      6 &      0 &  0 &    0 &   &  &  & 6 & 0  & 0 & 1     \\
             &              &      7 &      0 &  0 & 0 &    &  &  & 7 & 0  & 0& -1    \\
             &              &      8 &      0 &  0 &    0 &    &  &  & 8 & 0  & 0 & 1    \\

\end{tabular*}
\end{table}

\begin{table}[h]
\caption{Final six basis vectors ($\bfpsi$) for irrep $\Gamma_5$}
\begin{tabular*}{0.7\textwidth}{c@{\extracolsep{\fill}}cc|rrr|rrrr|rrr}
  $\Gamma$  &  $\bfpsi$  &  Er & \multicolumn{3}{c}{components} & & $\Gamma$  &  $\bfpsi$  &  Er & \multicolumn{3}{c}{ components}\\
      &      &             &$m_{a}$ & $m_{b}$ & $m_{c}$ &       &      &     &        &$m_{a}$ & $m_{b}$ & $m_{c}$  \\
\hline
$\Gamma_{5}$ & $\bfpsi_{19}$ &     1 &     0 &     0 &   0  & & & $\bfpsi_{22}$ & 1 & 1  & 0  &  0  \\
             &              &     2 &    0 &     0 &     0  &  &            &    & 2 & -1 & 0 & 0   \\
             &              &      3 &    0 &      0 &      0 &    &      &       & 3 & -1 & 0 & 0     \\
             &              &      4 &    0 &      0 &      0 &    &      &      & 4 & 1 & 0 & 0     \\
             &              &      5 &      0 &     -1 &      0 &    &        &    & 5 & 0 & 0 & 0   \\
             &              &      6 &      0 &    -1 &      0 &  & &             & 6 & 0 & 0 & 0        \\
             &              &      7 &      0 &    -1 &      0 & &   &            & 7 & 0 & 0 & 0      \\
             &              &      8 &      0 &    -1&      0 & &  &             & 8 & 0 & 0 & 0     \\
             & $\bfpsi_{20}$ &      1 &      0 &    0 &    0 & &  & $\bfpsi_{23}$ & 1 & 0  & 1  &  0     \\
             &              &      2 &      0 &    0 &      0 & & &              & 2 & 0 & 1 & 0     \\
             &              &      3 &      0 &   0 &      0 & &   &            & 3 & 0 & 1 & 0     \\
             &              &      4 &      0 &   0 &      0 & &  &             & 4 & 0 & 1 & 0    \\
             &              &      5 &     -1 &    0 &      0 & &  &              & 5 & 0 & 0 & 0         \\
             &              &      6 &    1 &    0 &      0 & &   &             & 6 & 0 & 0 & 0         \\
             &              &      7 &     1 &      0 &      0 & &    &          & 7 & 0 & 0 & 0      \\
             &              &      8 &    -1 &      0 &      0 & &    &          & 8 & 0 & 0 & 0       \\
             & $\bfpsi_{21}$ &      1 &   0 &   0 &  0 &   & & $\bfpsi_{24}$ & 1 & 0  & 0  & 1         \\
             &              &      2 &      0 &   0 &  0 &  &  &  & 2 & 0  & 0  & 1      \\
             &              &      3 &      0 &   0 &  0 &   &  &  & 3 & 0  & 0  & -1      \\
             &              &      4 &      0 &   0 &   0 &  &  &  & 4 & 0  & 0  & -1      \\
             &              &      5 &      0 &   0 &  -1 &   &  &  & 5 & 0  & 0 & 0    \\
             &              &      6 &      0 &  0 &    -1 &   &  &  & 6 & 0  & 0 & 0     \\
             &              &      7 &      0 &  0 & 1 &    &  &  & 7 & 0  & 0& 0   \\
             &              &      8 &      0 &  0 &  1 &    &  &  & 8 & 0  & 0 & 0    \\

\end{tabular*}
\end{table}

\clearpage
%